\documentclass[twocolumn,secnumarabic,amssymb,superscriptaddress, footinbib, aps, prx,reprint]{revtex4-1}
\usepackage{pdfpages}
\usepackage{xcolor}
\usepackage{float}
\usepackage{lipsum}
\usepackage{graphicx}% Include figure files
\usepackage{dcolumn}% Align table columns on decimal point
\usepackage{bm}% bold math
\usepackage[normalem]{ulem}
\usepackage{soul}
\usepackage{dblfloatfix}
\usepackage{amsmath}
\usepackage{color}  % Allows \textcolor{blue}{Hello world}
\usepackage{upgreek}

\makeatletter
\AtBeginDocument{\let\LS@rot\@undefined}
\makeatother

\begin{document}

\title{Chiral transport of hot carriers in graphene in the quantum Hall regime}

\author{Bin Cao}
%\thanks{These two authors contributed equally}
\affiliation{Joint Quantum Institute, NIST/University of Maryland, College Park, Maryland 20742, USA}

\author{Tobias Grass}
%\thanks{These two authors contributed equally}
\affiliation{ICFO-Institut de Ciencies Fotoniques, The Barcelona Institute of Science and Technology, Castelldefels (Barcelona) 08860, Spain}

\author{Olivier Gazzano}
\affiliation{Joint Quantum Institute, NIST/University of Maryland, College Park, Maryland 20742, USA}

\author{Kishan Ashokbhai Patel}
\affiliation{L-NESS, Department of Physics, Politecnico di Milano, Via Anzani 42, 22100 Como, Italy}

\author{Jiuning Hu}
\affiliation{NIST, Gaithersburg, Maryland 20878}

\author{Markus M\"{u}ller}
\affiliation{Joint Quantum Institute, NIST/University of Maryland, College Park, Maryland 20742, USA}

\author{Tobias Huber}
\affiliation{Joint Quantum Institute, NIST/University of Maryland, College Park, Maryland 20742, USA}

\author{Luca Anzi}
\affiliation{L-NESS, Department of Physics, Politecnico di Milano, Via Anzani 42, 22100 Como, Italy}

\author{Kenji Watanabe}
\affiliation{National Institute for Materials Science, 1-1 Namiki, 305-0044 Tsukuba, Japan}

\author{Takashi Taniguchi}
\affiliation{National Institute for Materials Science, 1-1 Namiki, 305-0044 Tsukuba, Japan}

\author{David Newell}
\affiliation{NIST, Gaithersburg, Maryland 20878}

\author{Michael Gullans}
\affiliation{Joint Center for Quantum Information and Computer Science,
NIST/University of Maryland, College Park, Maryland 20742 USA}

\author{Roman Sordan}
\affiliation{L-NESS, Department of Physics, Politecnico di Milano, Via Anzani 42, 22100 Como, Italy}

\author{Mohammad Hafezi}
%\email{hafezi@umd.edu}
\affiliation{Joint Quantum Institute, NIST/University of Maryland, College Park, Maryland 20742, USA}
\affiliation{IREAP, University of Maryland,
College Park, Maryland 20742, USA}

\author{Glenn Solomon}
%\email{gsolomon@umd.edu}
\affiliation{Joint Quantum Institute, NIST/University of Maryland, College Park, Maryland 20742, USA}

\date{\today}

\begin{abstract}
Photocurrent (PC) measurements can reveal the relaxation dynamics of photo-excited hot carriers beyond the linear response of conventional transport experiments, a regime important for carrier multiplication. In graphene subject to a magnetic field, PC measurements are able to probe the existence of Landau levels with different edge chiralities which is exclusive to relativistic electron systems. Here, we report the accurate measurement of PC in graphene in the quantum Hall regime. Prominent PC oscillations as a function of gate voltage on samples' edges are observed. These oscillation amplitudes form an envelope which depends on the strength of the magnetic field, as does the PCs' power dependence and their saturation behavior. We explain these experimental observations through a model using optical Bloch equations, incorporating relaxations through acoustic-, optical- phonons and Coulomb interactions. The simulated PC agrees with our experimental results, leading to a unified understanding of the chiral PC in graphene at various magnetic field strengths, and providing hints for the occurrence of a sizable carrier multiplication.

\end{abstract}

\maketitle

\section{Introduction} 
Chiral transport is a signature feature in many topological systems~\cite{hasan2010colloquium,qi2011topological} and results from restrictions in the motion of a particle at the edge of a gapped two-dimensional system to a single direction~\cite{halperin1982quantized,hasan2010colloquium,jackiw1976solitons}. In quantum Hall systems, the edge chirality of the charge carriers is determined by the direction of the magnetic field~\cite{halperin1982quantized,girvin1999quantum,patlatiuk2018evolution}. A special case is the quantum Hall effect in graphene where the relativistic nature of the electrons leads to particle-hole symmetry, and the absence of an intrinsic bandgap allows for both the relaxation of carriers and the tunability of the Fermi level across the Dirac point. Importantly, carriers within LLs across the Dirac point have opposite edge chirality~\cite{williams2007quantum,abanin2007quantized,nazin2010visualization,queisser2013strong}, with an exception of the zeroth LL ($\mathrm{LL}_0$)~\cite{abanin2007charge,abanin2007dissipative,kim2021edge}. Therefore, in optical experiments which excite electrons from LLs far below the Dirac point to LLs high above, an interesting interplay occurs between carriers of different types (electrons and holes) within LLs of potentionally different edge chiralities~\cite{nazin2010visualization,sonntag2017giant}. In particular, the edge chirality of a carrier may change during the relaxation process depending on the energy of the Fermi level with respect to the Dirac point, and in addition, the relaxation rates of electrons and holes may change and become unequal depending on the position of the Fermi level within a given LL. The result is a rich variety of photocurrent (PC) patterns whose detailed measurements are reported here, along with a microscopic modelling and an intuitive picture that together form a cohesive explanation of the observed behavior.

Qualitative studies of the PC in the quantum Hall regime have been reported~\cite{cao2016photo,nazin2010visualization,sonntag2017giant,wu2016multiple,gazzano2019observation}, giving explanations of the PC with heating effects~\cite{cao2016photo,wu2016multiple,wu2016device} or hot carrier relaxation with possible contributions from carrier multiplications (CM)~\cite{nazin2010visualization,wendler2014carrier,wendler2015ultrafast,song2013photoexcited}. However, a unified picture of the PC is still missing. Despite many attempts with optical methods~\cite{tielrooij2013photoexcitation,mittendorff2015carrier,plotzing2014experimental}, definitive evidence of CM in a Landau quantized monolayer graphene using electric measurements is still ambiguous~\cite{massicotte2021hot,gierz2013snapshots,wu2016multiple}. Experimentally, PC measurements from graphene in the quantum Hall regime are commonly obscured by the sample's substantially varying impedance when sweeping the Fermi level through many LLs~\footnote{See section \ref{results} and Appendix \ref{Section:Transport_and_PC_measurements}\ref{Subsection:TIA} for detailed discussions. Also see Ref.~\cite{gazzano2019observation}}. In addition, spatial scans of PC can be correlated with inhomegenous doping and charge puddles~\cite{nazin2010visualization}. These make the analysis of the PC mechanism even more challenging. 

Here, in order to measure the PC accurately, we use a trans-impedance-amplifier (TIA) which allows an accurate and precise measurement of PC, independent of the sample's varying impedance~\cite{horowitz2002art,mciver2020light,donati2021photodetectors}, while simultaneously keeping other parameters strictly controlled. At fixed magnetic fields, we observe prominent PC oscillations with opposite polarities on each of the two edges of the sample, as the Fermi level is swept through LLs. The envelopes of these oscillations are mapped as a function of magnetic field strength. In addition, we find the PC saturates with optical intensity and the critical power of this saturation changes as the Fermi level sweeps through LLs. To explain the dynamics, we model the observed PC using optical Bloch equations. Specifically, we model the relaxation of excited carriers considering acoustic, optical phonon and Coulomb scatterings. The PC is calculated based on hot carrier populations and their edge chiralities (group velocities) arising from energy dispersions on the edges. The simulation indicates that the PC oscillation is due to the alternating balance of available number of states for electrons and holes as we sweep the Fermi level through quantized LLs. As carriers relax across the Dirac point, the PC contributions from electrons and holes change from constructive to destructive, explaining the observed pattern of critical saturation powers. In addition, we find that the simulated PC matches the measurement at high magnetic fields only when Coulomb scattering is included, revealing evidence for CM. The paper is structured as the following: we first report the experimental observations of PC in section~\ref{results}. The model used to explain the measured data is detailed in section~\ref{model}. Finally, the measurements and the simulations are compared and conclusions are made in section~\ref{discussion}.

\onecolumngrid

\begin{figure}[H]
\centering
\includegraphics[scale=1]{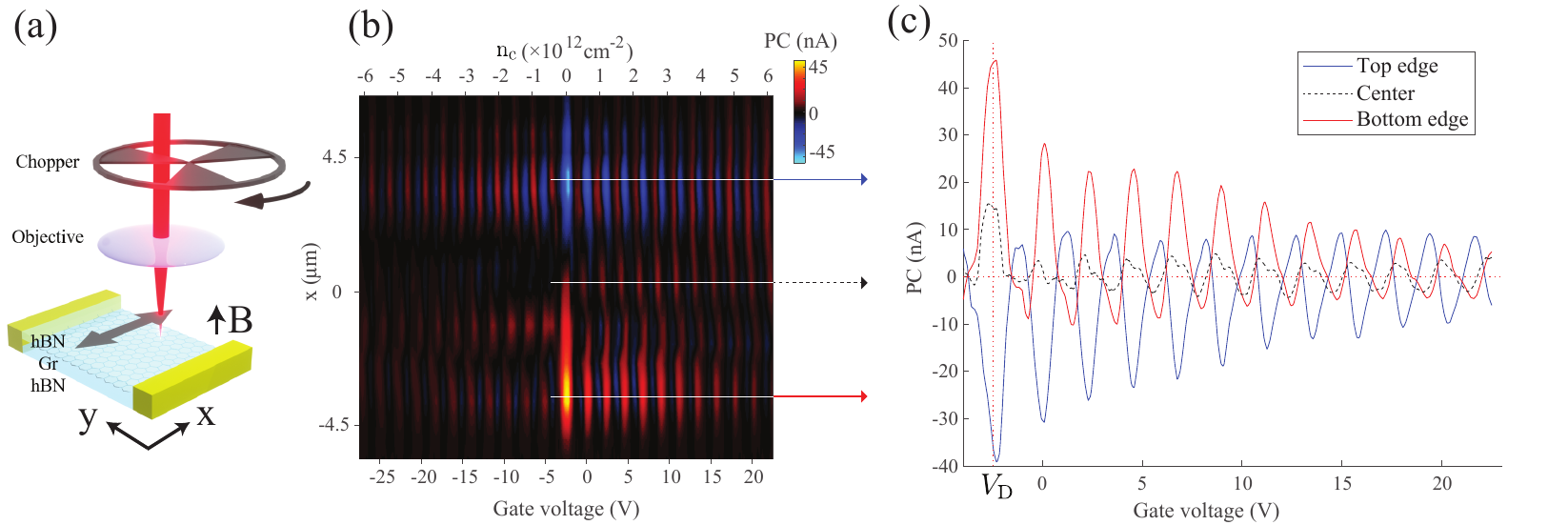}
\caption{(a) Schematics of the setup. We measure the PC along the center-line of the sample in the $x$ direction (grey arrow) as a function of position and gate voltage. The PC is measured at +$B$ and -$B$. The PCs of two antisymmetric fields are subtracted to isolate the $B$-dependent part of the PC and it is plotted in (b) for $B=4.5$ T, while the much weaker $B$-independent PC is shown in Fig.~(\ref{SI_Chip4_sample2065_4T_sum}) in the Appendix. In (b), prominent PC oscillations are observed on the edges (located at $x=\pm 4.5~\upmu m$) while the PC is minimal in the bulk (located around $x=0~\upmu m$). This indicates the PC is predominately attributed to the edge states. We plot the cuts of PC on the edges in (c). The zeros of the PC at high LLs match with the even-fillings (ignoring spin) of the LLs (systematically shown in Fig.~\ref{Fig2}(a)). PC oscillations on the two edges have opposite polarities, again indicating the PC is related to the chiral transport of carriers on the edges.}
\label{Fig1}
\end{figure}

\twocolumngrid

\section{Experimental observations}\label{results}
The experimental setup is shown in Fig.~(\ref{Fig1})(a). We focus a stabilized weak continuous-wave (CW) laser ($<10~\upmu W$, $\lambda=940~\rm{nm}$) on a two-terminal square sample with dimensions ($\sim 10~\upmu \rm{m}$) much larger than the laser spot ($\sim 1~\upmu \rm{m}$). The sample is at a temperature $\sim$ 4~K and has a mobility of 13100~$\mathrm{cm^2/(Vs)}$. The laser is chopped by a mechanical chopper with a frequency $\sim 300$~Hz. We move the laser spot position on the sample, and record the PC as a function of the gate voltage through two electrical contacts. The sample is not biased by any external voltage. The PC is measured via a home-made TIA which converts current to voltage independent of the sample's substantially varying impedance when sweeping the Fermi level through many LLs. This guarantees accurate measurements of the PC, including the corresponding current amplitude envelopes (see Appendix section~\ref{Section:Transport_and_PC_measurements}\ref{Subsection:TIA} for discussions) as well as minimizes the Johnson–Nyquist noise from the sample~\cite{donati2021photodetectors}. The voltage output of the TIA is measured with a lock-in amplifier which is frequency-locked with the chopped laser. Our measurements are repeated on two samples (see Appendix).

\begin{figure}[H]
\centering
\includegraphics[scale=1.5]{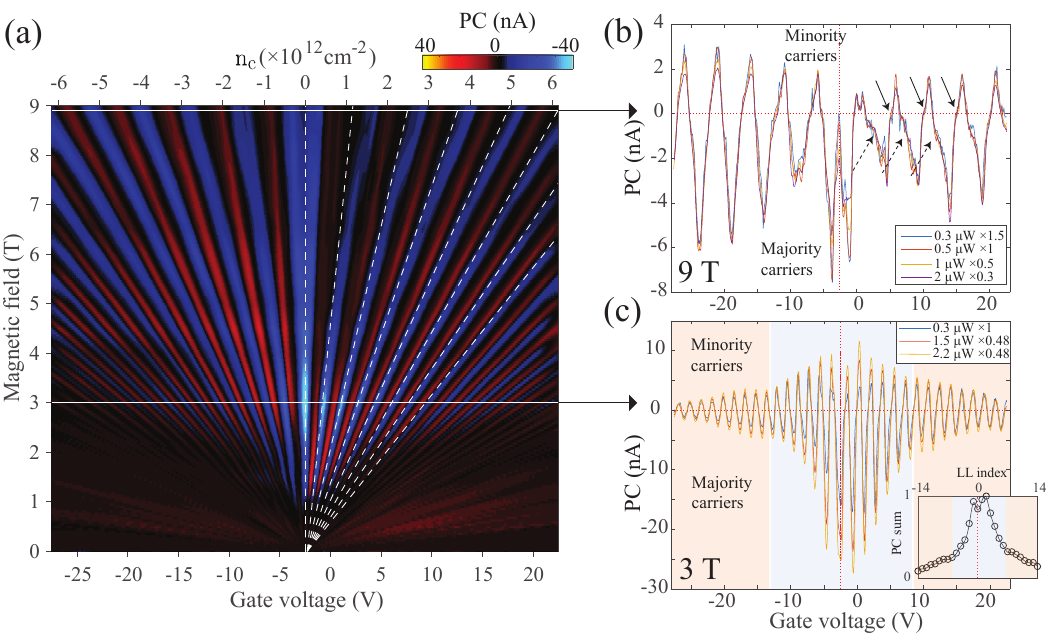}
\caption{The PC on the top sample edge as a function of field strength and gate voltage is plotted in (a) showing the Landau fan for the PC. The excitation power used is $1~\upmu W$. Data for $0.3~\upmu W$ and $2~\upmu W$ is shown in Fig.~(\ref{SI_PC_LL_fan_4_17_2020_chip4sample2065_0.3uW_and_2uW}) in the Appendix. The dashed white lines are even-fillings of LLs (ignoring spin) extracted from the transport fan (see Fig.~(\ref{SI_Transport_LL_fan_two_samples}) in the Appendix). As $E_\mathrm{F}$ is scanned, each LL (except the 0th) gives rise to a positive and a negative PC peak. In (b), cuts of PC for various pump intensity at a high field of 9 T are shown, and are scaled by factors in the legend. The shape of the oscillations shows that polarity changes at even-fillings (dashed black arrows) are smoother than that at odd-fillings (solid black arrows). We also see a prominent dip at the Dirac point which is due to efficient carrier relaxation when $E_\mathrm{F}=0$. In (c), cuts of the PC (scaled vertically by factors shown in the legend) for various pump intensities at a low field of 3 T are plotted. Based on measurements and simulations, the side with a more substantial envelope (the negative side) is attributed to the majority carriers and the other (the positive side) to the minority carriers, as marked in (b) and (c). The scaled PC cuts overlap well for high LLs but not near the Dirac point, indicating inhomogeneous PC power dependence with respect to LLs. In addition, we sum over the absolute values of the two PC peaks originating from the same LL and plot as a function of LL index in the inset of (c). The plot shows two regimes shaded in yellow and blue where the blue regime have larger slopes than the yellow, indicating different mechanisms, consistent with our model. Error bars are smaller than markers. Measurements on the other sample, showing same behaviors, are given in the Appendix.}
\label{Fig2}
\end{figure}

First, we spatially map the PC on the sample at a fixed magnetic field, as illustrated in Fig.~\ref{Fig1}(a) with experimental data in Fig.~\ref{Fig1}(b) and Fig.~\ref{Fig2}(b,c). Specifically, we stabilize the laser spot at a fixed position on the centerline of the sample and record the PC as a function of the gate voltage. Then we scan the laser spot position from edge to edge along the perpendicular centerline to obtain a two-dimensional plot (v.s. x and gate voltage). We find that the strength and polarity of the PC depend on: the position of the laser, as we scan the laser spot across the sample, along the centerline, as shown in Fig.~\ref{Fig1}(a); and the gate voltage, controlling the Fermi level in the sample. We have determined that the PC is independent on the laser polarization. Thus we ignore the spin degree of freedom and define even-fillings as $E_\mathrm{F}$ in the middle of LLs, whereas odd-fillings as in the middle of LL gaps. Next, we reverse the magnetic field, and we find that the PC signals are approximately reversed as well, especially when the laser spot is close to a sample edge. Therefore, we subtract the scans for $+B$ and $-B$ to separate the $B$-dependent and $B$-independent parts of the PC. The $B$-dependent PC is plotted in Fig.~\ref{Fig1}(b) as a function of laser spot position and the gate voltage. The much weaker $B$-independent PC, caused by direct diffusion~\cite{song2014shockley}, is shown in Fig.~(\ref{SI_Chip4_sample2065_4T_sum}) in the Appendix. In Fig.~\ref{Fig1}(c), we further visualize our data by showing cuts of the $B$-dependent PC as a function of gate voltage at three different positions on the sample: top edge, bottom edge, center. These plots highlight various features of the PC measurement: First, the $B$-dependent PC is strongly enhanced at the edges. Second, the PC oscillates with the gate voltage, and is strongest at the edges. Third, opposite edges give PC oscillations with opposite polarities. Fourth, PC peak values decrease as the backgate voltage increases away from the Dirac point.

\onecolumngrid

\twocolumngrid

We repeat our measurement on the top edge at different magnetic field strength up to 9~T. We find that the PC exhibits a Landau fan, shown in Fig.~\ref{Fig2}(a). Notably, this fan closely resembles the fan seen in conventional transport measurements (transport measurements are plotted in the Appendix). Additional features of the PC measurement are revealed by comparing PC slices at different fields, \emph{e.g.}, at 9~T and 3~T as shown in Fig.~\ref{Fig2}(b) and Fig.~\ref{Fig2}(c) for various laser intensities, where each PC slice is scaled vertically such that all the slices overlap at large backgate voltages. At low field (Fig.~\ref{Fig2}(c)), the PC peaks form an envelope with larger PC values closer to the Dirac point, while the envelop diminishes at high field (Fig.~\ref{Fig2}(b)). The envelope at low field can be separated into two regimes, with different slopes of PC as a function of gate voltage. This is highlighted in the inset of Fig.~\ref{Fig2}(c), where we sum over the absolute value of two PC peaks (positive and negative) arising from the same LL and plot as a function of LL index. The positive PC values are due to minority carriers (i.e. electrons in the valence band or holes in the conduction band) while the negative PC values are due to majority carriers (i.e. holes in the valence band or electrons in the conduction band), indicated on the left of Fig.~\ref{Fig2}(c). This is discussed below.

\begin{figure}[H]
\centering
\includegraphics[scale=1.2]{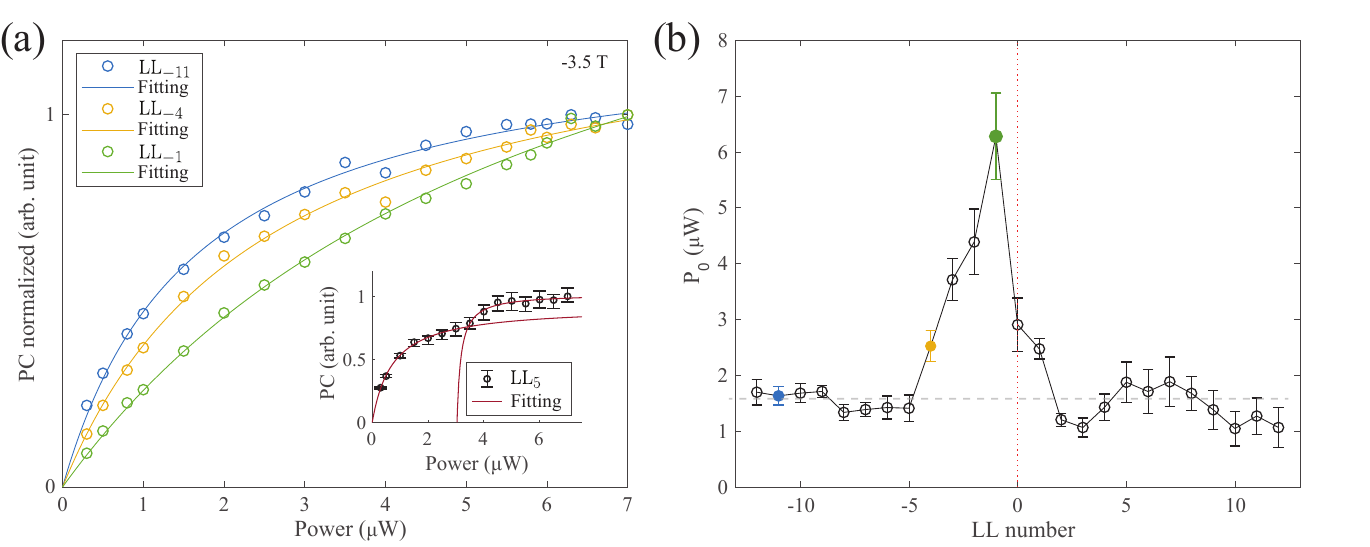}
\caption{Normalized PC at various LLs on the majority carriers side, as a function of optical power, at 3.5 T is plotted in (a). In (a), we observe little sign of PC saturation for $\mathrm{LL}_{-1}$, while for LLs with larger indices ($\mathrm{LL}_{-4}$ and $\mathrm{LL}_{-11}$), the PC is saturated more easily. The largest contribution to this effect is the carriers (electrons and holes) from above or below the Dirac point: At high LLs, the PC is proportional to the \textit{subtraction} of hot electrons and holes populations, which is limited by the relaxation rates difference (relaxation bottleneck); In contrast, for the low LLs, the PC is given as a \textit{sum} over the populations of the two. In contrast, at a higher field of 9 T, we see little sign of saturation (see Fig.~(\ref{SI_Saturation_3_7_2020_chip1_sample4_-9T}) in the Appendix). The error bars in (a) are smaller than the markers. The fitted corresponding saturation powers $P_0$ of data in (a) are plotted in (b). For high LLs, $P_0$ remains flat (gray dashed line as a guide for the eyes) which is a result of the relaxation bottlenecks for $E_{\mathrm{F}}$ at different high LLs saturated at the same power. The error bars represent 95\% confidence of fittings. In addition, we also observe a double-bent saturation behavior, as shown in the inset of (a), observed in a slightly different position. This is discussed in the main text.}
\label{Fig3}
\end{figure}

Finally, we study the PC amplitude dependence on the laser intensity. By comparing the maximum PC amplitudes (negative PC values in Fig.~\ref{Fig2}(c), which is due to majority carriers) at various LLs with different laser intensity, shown in Fig.~\ref{Fig3}(a), we find that when the magnetic field is low, the PC exhibits only weak saturation when the Fermi level is near the the Dirac point. But when the Fermi level is away from the Dirac point (above $\mathrm{LL}_{\pm5}$), the PC saturates at a constant laser power as shown in Fig.~(\ref{Fig3})(b). The saturation powers $P_0$ are fitted with the saturable absorber model~\cite{winzer2017unconventional,shen1984principles},
\begin{equation}
    J \propto \frac{P/P_0}{1+P/P_0}.
\end{equation}
The extracted $P_0$ corresponding to $\mathrm{LL}_{-12}$ to $\mathrm{LL}_{+12}$ is plotted in Fig.~\ref{Fig3}(b).

\onecolumngrid

\twocolumngrid

\section{Theoretical model}\label{model}
In order to model our data, we first recall that quantum Hall systems are insulating in the bulk, but they have $B$-dependent chiral transport on the edge when $E_\mathrm{F}$ is within a LL gap. Therefore, we assume that the $B$-dependent PC is given by intralevel currents from the system edge. We denote the carrier population by $n_{\nu,k}$, where $\nu$ is the LL index, and $k$ is the wave number. The population shall refer to electrons above the Fermi level, and to holes below the Fermi level. The PC generated by excited carriers is given by $J \sim \sum_{\nu} \int dk q_\nu v_{\nu,k} n_{\nu,k}$, where $v_{\nu,k}$ is the group velocity of the given level, and $q_\nu$ takes into account the charge of the carrier: $q_\nu=1$ when $\nu$ is below the Fermi level; $q_\nu=-1$ when $\nu$ is above the Fermi level. Specially, $q_\nu=0$ when $\nu$ is at the Fermi level (i.e. for carriers within the LL where the Fermi level is), to account for the existence of both electrons and holes in the vicinity of the Fermi level and the divergent acoustic phonon couple strength $\propto 1/\omega$ in the limit of small frequencies~\cite{wendler2015ultrafast}. To simplify the situation, we assume a constant group velocity $v_{\nu,k}=v_{\nu}$, whose precise value depends on the position of the laser spot on the sample. With this simplification, the current will be determined by average population of the LLs, $n_\nu \sim \int dk n_{\nu,k}$. Due to opposite velocities on the two edges, $v_{\nu}$ changes sign as the laser is moved from one edge to the opposite edge, explaining the PC polarity shift as a function of position in Fig.~\ref{Fig1}(b).

\begin{figure}[H]
\centering
\includegraphics[scale=1]{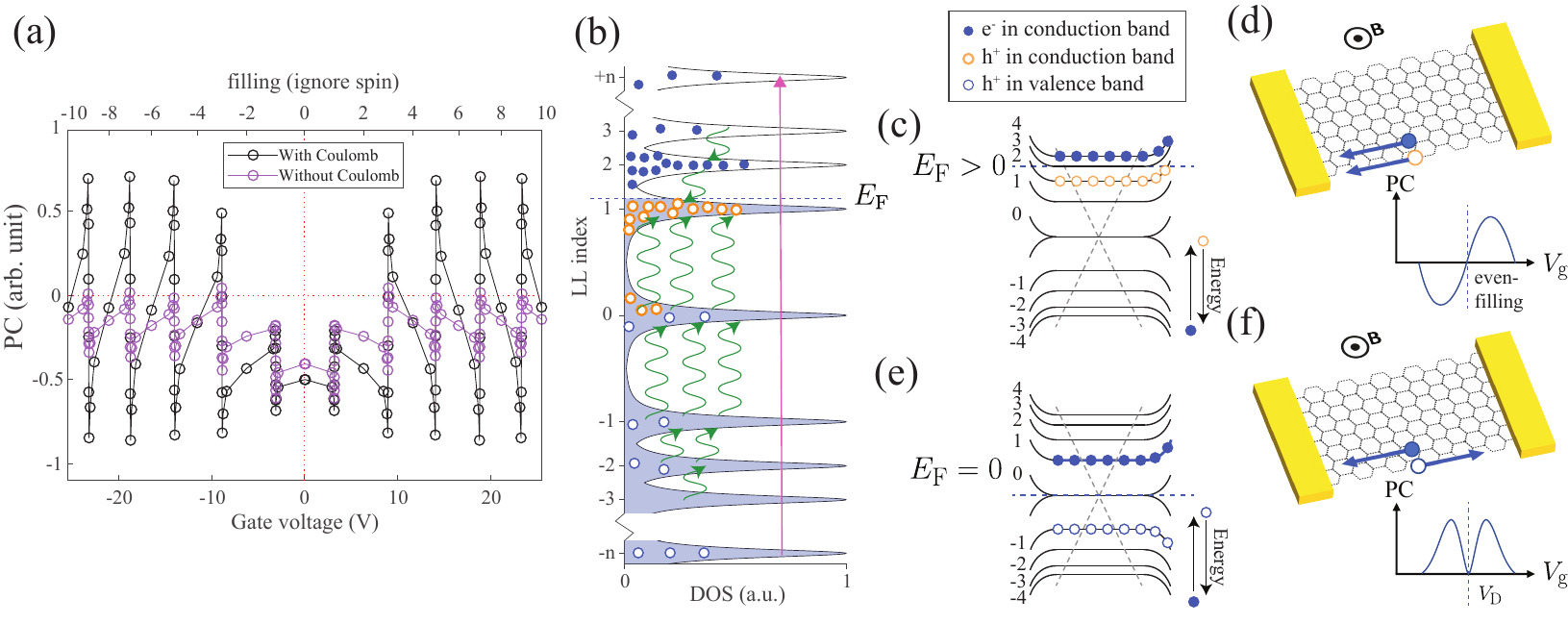}
\caption{The simulated PC, with and without Coulomb interactions, as a function of gate voltage is plotted in (a). The simulated PC with Coulomb interactions matches the measured PC at 9 T shown in Fig.~(\ref{Fig2})(b) whereas the simulation without Coulomb interactions does not, showing the importance of contributions from Coulomb interactions in the PC. In particular, without Coulomb interaction, the simulated PC decreases with gate voltage away from the Dirac point, whereas the opposite trend appears for the case with Coulomb which matches with the observation in Fig.~(\ref{Fig2})(b). In addition, the simulations with Coulomb interactions have asymmetries between the positive and negative PC peaks away from the Dirac point that match well the data represented in Fig.~(\ref{Fig2})(b), whereas the simulations without Coulomb interactions do not. Moreover, the asymmetric LL PC peak shape also matches the observations (see arrows in Fig.~(\ref{Fig2})(b)). In (b), we illustrate the carrier relaxation across LLs. Depending on $E_\mathrm{F}$, excited electrons and holes may relax to the same side of the Dirac point. These electrons and holes share the same edge chirality (\emph{e.g.} the holes above the Dirac point (orange hollow circles) and the electrons (blue solid dots)) and therefore give rise to the destructive part of the PC. To further illustrate this, we show in (c) that when electrons and holes are on the same side of the Dirac point,  edge chiralities are the same as shown in (d) giving a destructive PC; in (e) when electrons and holes are on different sides of the Dirac point, edge chiralities are opposite as shown in (f) giving a constructive PC.}
\label{Fig4}
\end{figure}

At a fixed laser position, due to opposite edge chirality of LLs $\nu$ and $-\nu$,  the velocities take opposite values: $v_{\nu}=-v_{-\nu}$, as illustrated in Fig.~\ref{Fig4}(e) and (f). In view of the otherwise identical transport properties in LLs, we take the group velocity to be a constant, \emph{i.e.}, $|v_{\nu}|=v$~\cite{sonntag2018impact}. An exception from this occurs when $\nu=0$ ($\mathrm{LL}_0$), where the lack of well-defined edge chirality yields $v_0=0$ due to nonequilibrium distributions~\cite{abanin2007charge,abanin2007dissipative}. With these simplifications, the PC signal is \[J\sim v \sum_{\nu\neq 0} q_\nu {\rm sign}(\nu) n_{\nu}.\] This is similar to the Shockley-Ramo theorem~\cite{shockley1938currents,ramo1939currents} except that here the group velocities are given by the edge chirality rather than an electric field. 

To determine the average LL population $n_\nu$, we employ the optical Bloch equations, as described in details in Appendices \ref{Section:Quantitative_model} and \ref{Section:Pseudopotentials}. They determine the steady state in the presence of optical excitation and different relaxation channels (acoustic phonons, optical phonons, 
Coulomb scatterings). The result of our PC simulation is shown in Fig.~\ref{Fig4}(a). In particular, our PC model reproduces the PC oscillations as a function of gate voltage seen in the experiment. With all relaxation channels considered, the envelope of the simulated current matches with the experimentally obtained envelope in the high field regime. This is understood as certain assumptions in our simulation implicitly assume a strong magnetic field. Specifically, we neglected mixing between LLs, and the number of LLs in the simulation was limited to 21, which is close to the resolvable number of LLs~\footnote{The number of resolvable LLs is $\sim$ 29 and this number does not change with B~\cite{orlita2011carrier}. We limit to 21 due to computational cost }.

\onecolumngrid

\twocolumngrid

\section{Discussion}\label{discussion}
The $B$-dependent part of PC is dominated by carriers at the edge, see Fig.~\ref{Fig1}(b). If all carriers on the edge contribute equally to the PC, the measured PC should be strictly zero. To explain the prominent PC oscillations on the edge, we develop an effective model where the PC is due to the hot carriers whereas the carriers near $E_\mathrm{F}$ has little contribution due to the divergent acoustic phonon coupling strength between electrons and holes at $E_\mathrm{F}$ which has been elaborated above.  

The polarity of edge current has two contributions. First, it depends on the chirality of the particular edge state~\cite{nazin2010visualization}, which is incorporated in our theoretical model through the opposite group velocities for LLs above and below the Dirac point, as shown in Fig.~\ref{Fig4}(c-f). Second, the current polarity also depends on the charge of the hot carrier. Thus, if the Fermi level is sufficiently far away from the Dirac point, all hot carriers in the vicinity of the Fermi level will be on the same side of the Dirac point and contribute to the PC with the same edge chirality~\cite{lee2017inducing,shalom2016quantum}. However, electrons and holes contribute with opposite charges, therefore, in this case the PC signal reflects a mismatch (a subtraction) between the number of hot electron carriers and hole carriers. The magnitude of this mismatch crucially depends on the position of the Fermi energy within the LL. More specifically, as we sweep $E_{\mathrm{F}}$ through a LL, the available number of states for electrons and holes alternates (illustrated in Fig.~\ref{Fig4}(b)), generating unbalanced carrier numbers in accordance with Fermi's golden rule. This mechanism pictorially explains the PC oscillations, cf. also ref.~\cite{nazin2010visualization}, which are also reproduced by our microscopic quantitative model invoking the relaxation of carriers, with edge chirality after photoexcitation. Details of the oscillation asymmetries, oscillation envelop structure and the shape of oscillations are discussed below in the context of our model.

In the low field case --- see Fig.~\ref{Fig2}(c) for instance, asymmetries in the absolute values of the maximum PC between the majority and minority side are present at lower backgate voltage, i.e., near the Dirac point. As the Fermi level nears the Dirac point, the occupation of chiral edge states both above and below the Dirac point must be considered. Without considering carrier type, edge states on opposite sides of the Dirac have opposite chirality. Thus, if a carrier population extends across the Dirac point, populations of the same carrier type on each side of the Dirac point contribute to opposite PC directions. Therefore, hot carrier populations above the Dirac point will add constructively with opposite carrier types below the Dirac point and destructively with opposite carrier types on the same side of the Dirac point~\cite{nazin2010visualization}, leading to asymmetries in the maxima of the absolute values of PC oscillations (Fig.~\ref{Fig2}(c)). In both the experimental data, Figs.~\ref{Fig1}(c) and \ref{Fig2}, and in the simulation data, Fig.~\ref{Fig4}(a), we observe this asymmetry between positive and negative PC peaks. 

This asymmetry manifests differently in high and low magnetic fields. In the high field case, due to the larger LL gaps, there are less LLs for relaxation compared to the low field case. As a result, most of the relevant LLs have well-defined edge chiralities. Therefore, the asymmetry manifests as a constant offset to PC for all LLs, which is shown in both the simulation (Fig.~\ref{Fig4}(a)) and measurements (Fig.~\ref{Fig2}(b)). In the low magnetic field case, the excitation energy corresponds to LLs with much larger index. These high LLs are merged together (measurements shown in Fig.~(\ref{SI_Wavelength_dependence_chip4_sample2065_870_940nm_3d5T_and_9T}) in the Appendix) and thus they do not have well-defined edge chirality due to the lack of gaps~\cite{hasan2010colloquium}. Therefore, in this case, the distribution of hot carriers only affects the PC peaks when $E_{\mathrm{F}}$ is in the low LLs and form the envelope as seen in Fig.~\ref{Fig2}(c), and it is elaborated next.

The concept of opposite carrier types on opposite sides of the Dirac point adding constructively to the PC reaches an extreme when the Fermi level is within the zeroth LL in the low field, again as shown in Fig.~\ref{Fig2}(c). With the Fermi level at the Dirac point, electrons necessarily belong to LLs above the Dirac point, whereas holes necessarily belong to LLs below the Dirac point. Together with the opposite edge chirality of electrons and holes regardless of their LL index, this leads to a situation in which their PC contributions are always constructive~\cite{sonntag2017giant} and no polarity change can occur. Thus, in the low magnetic-field case, this situation gives rise to an envelop in the PC oscillation, in which the majority carrier side is peaking near the Dirac point. While when $E_{\mathrm{F}}$ moves away from the Dirac point, the constructive component is reduced as the distribution of hot carriers moves across the Dirac point. When $E_{\mathrm{F}}$ is far away from the Dirac point in this low magnetic field case, electrons and holes are mostly located on the same side of the Dirac point, so that electron and hole current contributions add destructively and the PC oscillations are almost symmetric. The two regimes: $E_{\mathrm{F}}$ near and away from the Dirac point, are indicated using different background shading colors in Fig.~\ref{Fig2}(c). In the inset in Fig.~\ref{Fig2}(c) we sum over the absolute values of PC peaks belonging to the same LL to highlight the different slopes of the PC amplitudes versus backgate voltage. The two regimes\textendash one close to and one away from the Dirac point\textendash are clear, reflecting our model.

When the field strength is high as shown in Fig.~\ref{Fig2}(b), an important caveat to our model is the fast relaxation of both types of carriers when $E_\mathrm{F}=0$. This results in a dip of the PC at the Dirac point. This dip in our model is not due to sublattice symmetry breaking from a Moir\'e pattern~\cite{hunt2013massive,yang2021experimental} or interaction-induced valley symmetry breaking of the $\mathrm{LL}_0$~\cite{zhang2006landau,young2012spin,amet2013insulating,kim2021edge}, since none of the corresponding features, such as the minibands and LL splittings, is observed in our transport measurements (see Fig.~(\ref{SI_Transport_LL_fan_two_samples}) in the Appendix). Furthermore, this dip is seen in the strong field data of Fig.~\ref{Fig2}(b) and the simulation data in Fig.~\ref{Fig4}(a).

%As $E_{\mathrm{F}}$ is moved further away from the Dirac point, close to the maximums of the gate voltage, the LLs starts to slowly overlap and the edge chirality is hampered by the back-scatterings. Thus, PC oscillations decay down.

Using the PC envelopes, we see evidence for carrier multiplication~\cite{wendler2014carrier,mittendorff2015carrier}. Specifically, in Fig.~(\ref{Fig2})(c), as we sweep $E_{\mathrm{F}}$ from high LLs to the Dirac point, the majority carrier PC peaks increase in amplitude, as a consequence of minority carriers starts to distribute across the Dirac point, which is discussed above. This alone is not an evidence of CM, but carrier distribution around the Dirac point. However, the minority carrier peaks grow as well and the only explanation of both the majority and minority carrier PC peaks increase in our model is that carriers are multiplied as they relax from high LLs to low LLs. This is also accompanied by simulations: by turning off the Coulomb interactions, simulations show much weaker PC and only with Coulomb scatterings, would the simulated envelope (Fig.~\ref{Fig4}(a)) match the measured one at high field (Fig.~\ref{Fig2}(b)).

The unique shape of the individual oscillations within a LL is well-represented in our theoretical model and is strikingly similar to our high-magnetic field experimental data. This asymmetry in the individual oscillation shape gives insight into the hot-carrier relaxation process. When $E_{\mathrm{F}}$ sweeps pass the even-filling point in a LL, hot carriers of one type slowly outnumber the other type, leading to a smooth PC polarity flip. In contrast, when the LL is gapped and $E_{\mathrm{F}}$ sweeps pass an odd-filling point in the LL, the number of available states for relaxation quenches. This creates a relaxation bottleneck for one carrier type, whereas relaxation from the other carrier type emerges, leading to a shape change in PC. This is seen in the experimental data of Fig.~\ref{Fig2}(b), where the oscillations are clearly not sinusoidal. At positive gate voltage, we note that positive PC peaks are not aligned exactly in the middle of two negative PC peaks. Instead, they are shifted towards the Dirac point, which makes the PC polarity changes at odd-fillings (marked by solid black arrows in Fig.~\ref{Fig2}(b))  more abrupt than the ones at even-fillings (marked by dashed black arrows in Fig.~\ref{Fig2}(b)). This feature is matched in our simulation, Fig.~\ref{Fig4}(a).

Experimental data related to PC saturation are shown in Fig.~(\ref{Fig3}), and can provide some additional insight to carrier distributions. We observe that the PC for low LLs does not saturate, whereas the high LL PC saturates much easier. This is consistent with our model since the PC for low LLs is mainly a sum of the electron and hole currents; whereas in the high LL regime the PC saturates much easier since it is the difference of electron and hole currents and thus bottlenecked by the difference of relaxation rates of electrons and holes~\cite{pogna2021hot,massicotte2021hot}. This is evidenced by the observed markedly different saturation powers, $P_0$ for low and high LL (Fig.~(\ref{Fig3})(b)) which supports the different mechanisms of the PC generation in these two regimes. In addition, in Fig.~(\ref{Fig2})(c), we see the scaled PCs measured as a function of different laser powers overlap well in the high LLs, but do not for the low ones, indicating inhomogeneous PC power dependence with respect to LLs.
    
We also see a peculiar double-bent saturation behavior which occurs predominately at high LLs, as shown in the inset of Fig.~(\ref{Fig3})(a). Similar behavior has been observed in a high power pulsed regime~\cite{winzer2017unconventional} which is attributed to the efficient Coulomb scattering processes at the pumped energies. However, we observe at much lower intensities here possibly due to the continuous pump we are using, which constantly excites carriers and thus enhances the Coulomb interactions. We note that the double-bent saturation is sensitive with position and the exact mechanism needs further study of the impacts from sample sizes and disorder strength, etc.

In conclusion, we accurately measure the PC in graphene in the quantum Hall regime by using a TIA. The improved accuracy, allows us to correlate particular experimental features with our theory and simulations to form a coherent model of PC in the quantum Hall system. In particular, the balance between hot electrons and hot holes oscillates with the alternating available number of states for electrons and holes near the Fermi level, as it is swept through LLs. The chiral current contributions from these hot carriers are determined by the edge chirality of carriers originated from the confinement dispersion, together with their charge. As a consequence of the flipped edge chirality across the Dirac point as well as a hot carrier distribution over LLs, PC shows different saturation behaviors when the Fermi level is close or away from the Dirac point. In addition, inclusion of the carrier multiplication and Coulomb interaction is imperative to explain the simultaneously growing majority and minority carrier PC peaks in the low field and the PC oscillation pattern in the high field.

Our work provides an unique study of carrier relaxations using a continuous excitation, which extends beyond the ultrafast regime studied in most pump-probe measurements. We believe our coherent model of PC from graphene in the quantum Hall regime will further the development of applications using graphene such as single-photon detection~\cite{walsh2021josephson,tielrooij2015generation,koppens2014photodetectors} and light-harvesting~\cite{gabor2011hot,ma2019giant}, as an electrically-measured CM has been long-coveted~\cite{massicotte2021hot}. The quantum Hall regime can be realized with a synthetic gauge field~\cite{kang2021pseudo}. Our findings will also facilitate exploring new physics such as studies of twistronics~\cite{han2021accurate}, topological properties of quantum devices~\cite{PhysRevB.103.L241301} and controlling the CM~\cite{but2019suppressed}. In addition, using PC measurements may provide a new angle for studying and distinguishing topological edge currents~\cite{abanin2011giant,gorbachev2014detecting} and the recently confirmed non-topological edge currents~\cite{uri2020nanoscale,aharon2021long}.

\acknowledgments
The authors acknowledge Thomas Murphy and Martin Mittendorff for carefully reading the manuscript and insightful discussions with Alessandro Restelli and John Lawall.

The work in Maryland was supported by Grants No. ARO W911NF2010232, No. ARL
W911NF1920181, and No. AFOSR FA95502010223, the
Simons Foundation, and the NSF funded PFC@JQI. T.G. acknowledges a fellowship granted by “la Caixa” Foundation (ID100010434, fellowship code LCF/BQ/PI19/11690013), as well as funding from Fundacio Privada Cellex, Fundacio Mir-Puig, Generalitat de Catalunya (AGAUR Grant No. 2017 SGR1341, CERCA program, QuantumCAT U16-011424, co-funded by ERDF Operational Program of Catalonia 2014-2020), Agencia Estatal de Investigacion (“Severo Ochoa” Center of Excellence CEX2019-000910-S, Plan Nacional FIDEUA PID2019-106901GB-I00/10.13039/501100011033, FPI), MINECO-EU QUANTERA MAQS (funded by State Research Agency (AEI) PCI2019-111828-2/10.13039/ 501100011033), EU Horizon 2020FET-OPEN OPTOLogic (Grant No 899794), ERC AdG NOQIA, and the National Science Centre, Poland-Symfonia Grant No. 2016/20/W/ST4/00314. K.A.B, L.A., and R.S. acknowledge the EU Horizon 2020 Graphene Flagship
Core 3 Grant No. 881603.

B.C. and T.G. contributed equally to this work.

\clearpage

\appendix
\section{Quantitative model}\label{Section:Quantitative_model}
For a theoretical description of the system dynamics, we employ optical Bloch equations for Landau quantized graphene \cite{wendler2014carrier, wendler2015ultrafast, PhysRevB.103.L241301}. The population of $\mathrm{LL}_i$ (averaged over all orbitals) is denoted by $\rho_i$, and the (averaged) polarization between two LLs is denoted by $P_{ij}$. Their equations of motion read:
\begin{align}
     \dot \rho_i &= 2{\rm sign}(i) {\rm Re}\left(\Omega_{ij} P_{ij}\right)  + S_i^{(\rm in)} (1-\rho_i) - S_i^{(\rm out)}\rho_i, \\
     \dot P_{ij} &=  - \Omega_{ij}^*(\rho_i-\rho_j) - \frac{\Gamma}{\hbar} P_{ij}.
\end{align}
Here, we choose the rotating frame of the light field, in which we further employ the rotating wave approximation. With this, the Rabi frequency $\Omega_{ij}$ takes non-zero values $\Omega_0$  only for two resonant pairs of Landau levels, $(i,j)=(n_{\rm min}, n_{\rm max}-1)$ and $(i,j)=(n_{\rm min}+1, n_{\rm max})$. For the modeling, we assume $n_{\rm min}=-10$ and $n_{\rm max}=10$. Choosing the vector potential at a constant value $A=5 \times 10^{-10}~{\rm Vs}{/\rm m}$, the Rabi frequency is given by $\Omega_0 = \frac{e}{m_e} M_{if} A = 200~{\rm GHz}$, with an optical matrix element between non-zero Landau levels $M_{if}= 2.25\times 10^9~{\rm m^{-1}}$, cf. ref. \cite{wendler2015ultrafast}. In the scattering rates, we include both Coulomb interactions and phonon emission: $S_i^{\rm (in)} = S_i^{\rm (in, Coul)} +S_i^{\rm (in,ph)}$ and $S_i^{\rm (out)} = S_i^{\rm (out, Coul)} +S_i^{\rm (out,ph)}$. 

Within second-order Markov-Born approximation, the Coulomb scattering terms are given by:
\begin{align}
    S_i^{\rm (in, Coul)} =& \frac{2\pi}{\hbar} \sum_{jkl} V_{ijkl} (4 V_{klij} - V_{lkij}) \times \nonumber \\ & (1-\rho_j)\rho_k \rho_l \Gamma(\Delta E_{ijkl}), \\
    S_i^{\rm (out, Coul)} =& \frac{2\pi}{\hbar} \sum_{jkl} V_{ijkl} (4 V_{klij} - V_{lkij}) \times \nonumber \\ & \rho_j (1-\rho_k)(1-\rho_l)  \Gamma(\Delta E_{ijkl}).
\end{align}
Here, $V_{ijkl}$ are the (averaged) Coulomb interaction matrix elements for a scattering process from LLs $k$ and $l$ into LLs $i$ and $j$. The function $\Gamma(E)$ introduces a step-function broadening of the LLs,
$\Gamma(E)= \frac{1}{\pi\Gamma_0} \theta(\Gamma_0 + E) \theta(\Gamma_0-E)$, where we have chosen $\Gamma_0=9~{\rm meV}$~\cite{wendler2014carrier}. The energy difference $\Delta E_{ijkl}$ refers to the difference $\epsilon_i +\epsilon_j-\epsilon_k -\epsilon_l$ of unbroadened LLs, with $\epsilon_i =  {\rm sign}(i)\sqrt{2|i|}\frac{\hbar V_{\rm F}}{l_B}$, where $v_{\rm F}= 10^6~{\rm m/s}$ and $l_B=25~{\rm nm}/\sqrt{B}$. The Coulomb interaction matrix elements in graphene can be expressed in terms of Coulomb matrix elements $\tilde V_{ijkl}$ of non-relativistic LLs:
\begin{align}
    V_{ijkl} =& c_i c_j c_k c_l ( \tilde V_{|i|,|j|,|k|,|l|} + \tilde V_{|i|-1,|j|,|k|,|l|-1} + \nonumber \\ & \tilde V_{|i|,|j|-1,|k|-1,|l|}+ \tilde V_{|i|-1,|j|-1,|k|-1,|l|-1} ),
\end{align}
where we define $C_i=1$ if $i=0$, and $c_i={\rm sign}(i)/\sqrt{2}$ elsewise, and $\tilde V_{ijkl}=0$ if any of the indices becomes negative. For positive indices, we employ the pseudopotential model, and define $\tilde V_{ijkl}$ as the dominant pseudopotential between the involved LLs, i.e., the pseudopotential which scatters a pair of particles in LLs $k$ and $l$ and at relative angular momentum $m=0$ into LLs $i$ and $j$ and relative angular momentum $m'=i+j-k-l$.
An explicit construction of the pseudopotential is appended in Appendix \ref{Section:Pseudopotentials}.

The phonon emission processes are described by the following scattering terms:
\begin{align}
    S_j^{\rm (in,ph)} =& \int dE \sum_{i\neq j} \Gamma(\Delta E_{ij}- \hbar \omega) g_0 f(E) \rho_i,
    \\
    S_j^{\rm (out,ph)} =& \int dE \sum_{i\neq j} \Gamma(\Delta E_{ji}- \hbar \omega) g_0 f(E) (1-\rho_i)
\end{align}
where we assume $g_0= 36~{\rm THz}$ for an overall coupling frequency, while the scalar function $f(E)$ takes into account an additional energy-dependence of the coupling. Specifically, we assume acoustic phonons in the energy range from 1 meV to 160 meV, with a
$1/E$ dependence, see also ref.~\cite{wendler2015ultrafast,malic2013graphene}. For concreteness, we assume $f(E)=4~\mathrm{meV}/E$. In the energy ranges from 170 meV to 175 meV and from 190 meV to 195 meV, optical phonons give rise to an approximately energy-independent coupling, for which we take $f=1$.

The equilibrium carrier populations are solved from the Bloch equations. The current for different Fermi level is calculated based on the model discussed in the main text. The Fermi level is converted to the gate voltage through a convolution with the Lorentzian-shaped density of states of LLs, with a width of $2\Gamma_0$.

\section{Pseudopotentials}\label{Section:Pseudopotentials}
In Fourier space, the Coulomb potential reads,
\begin{align}
    V(Q) = \frac{e^2}{8\pi^2 \epsilon_0 \epsilon_{\rm d} l_B}  \frac{1}{Q} \equiv V_0/(2\pi Q),
\end{align}
and the interaction between two electrons at positions ${\bf r}_1$ and ${\bf r}_2$ is expressed as
\begin{align}
V=\int d{\bf q} V(|{\bf q}|) e^{i {\bf q}\cdot ({\bf r}_1- {\bf r}_2)}.
\end{align}
Using this expression, the scattering of a pair of electrons from (non-relativistic) LLs $n_1$ and $n_2$ into LLs $n_3$ and $n_4$ is determined by $\langle m M, n_1 n_2 | \exp[i {\bf q}\cdot ({\bf r}_1- {\bf r}_2)] | \tilde m \tilde M, n_3 n_4 \rangle$. Here, $m,\tilde m$ is the relative angular momentum of the pair, and $M,\tilde M$ their center-of-mass angular momentum. Following Ref~\cite{macdonald1994introduction}, we replace the position operators ${\bf r}_i$ by raising operators $a_i^\dagger$ and $b_i^\dagger$, with $a_i^\dagger$ raising the LL quantum number of particle $i$ by one, and $b_i^\dagger$ raising the angular momentum quantum number of particle $i$ by one. Conveniently, we re-write the $b$ operators in the $b_{r,R}$ basis, $b_r=(b_1-b_2)/\sqrt{2}$ and $b_R=(b_1+b_2)\sqrt{2}$. We obtain
\begin{align}
e^{i {\bf q}\cdot ({\bf r}_1- {\bf r}_2)} =& e^{-Q^2/2} e^{i\bar q a_1^\dagger /\sqrt{2}}  e^{i q a_1 /\sqrt{2}} e^{-Q^2/2} \nonumber \\ \times& e^{-i\bar q a_2^\dagger /\sqrt{2}}  e^{-i q a_2 /\sqrt{2}}  e^{i q b_r^\dagger}  e^{i \bar q b_r},
\end{align}
where $Q=|{\bf q}|$ and $q=q_x-iq_y$. 
Since no operator acts on the $M$ quantum numbers, the potential has the form $\delta_{M,\tilde M}$. The whole expression reads
\begin{align}
 V_{m \tilde m M \tilde M}^{n_1 n_2 n_3 n_4} = & \delta_{M,\tilde M} \int d^2 {\bf q} V({\bf q}) e^{-Q^2}
 \langle m| e^{i q b_r^\dagger}  e^{i \bar q b_r} |\tilde m\rangle \times
 \nonumber \\ &
 \langle n_1| e^{i\bar q a_1^\dagger /\sqrt{2}}  e^{i q a_1 /\sqrt{2}} 
| n_3 \rangle \times
\nonumber \\ &
\langle n_2| e^{-i\bar q a_2^\dagger /\sqrt{2}}  e^{-i q a_2 /\sqrt{2}} | n_4\rangle.
 \end{align}
The terms in brakets are evaluated as
\begin{align}
 \langle n| e^{i\bar q a^\dagger /\sqrt{2}}  e^{i q a /\sqrt{2}} |\tilde n \rangle =& \left(\frac{\tilde n!}{n!} \right)^{1/2} \left(\frac{iq}{\sqrt{2}}\right)^{n-\tilde n}  L_{\tilde n}^{n-\tilde n}\left(\frac{Q^2}{2}\right),
\end{align}
and
\begin{align}
 \langle m| e^{i q b_r^\dagger}  e^{i \bar q b_r} | \tilde m\rangle =& \left(\frac{\tilde m!}{m!} \right)^{1/2} \left(iq\right)^{m-\tilde m}  L_{\tilde m}^{m-\tilde m}\left(Q^2\right),
\end{align}
with $L_n^\alpha(x)$ denoting the generalized Laguerre polynomials. For the scattering matrix elements used in the equations of motion, $\tilde V_{n_1n_2n_3n_4}$, we have used the dominant pseudopotential, $\tilde V_{n_1n_2n_3n_4} \equiv V_{m \tilde m M \tilde M}^{n_1 n_2 n_3 n_4}$ with $m=0$,  $\tilde m = n_1+n_2-n_3-n_4$, and arbitrary $M=\tilde M$.

\section{Sample fabrication}
Graphene was exfoliated from natural graphite crystals (HQ Graphene) and hBN was exfoliated from a synthetic crystal~\cite{watanabe04}. Both materials were exfoliated using the Scotch-tape method on two different Si substrates with a dry-grown 90-nm-thick SiO$_2$ layer on top. Monolayer graphene and 10-15 nm thick hBN were identified based on the color contrast on their respective substrates by an optical microscope. Thicknesses of each material were later confirmed by micro-Raman~\cite{ferrari06} and atomic force microscopy. Heterostructures of hBN/graphene/hBN were assembled by a hot pick-up method~\cite{pizzocchero16} on the same type of substrates. The backside of the substrates was metallized and used as a back gate.

Assembled heterostructures were shaped into squares of different sizes, ranging from 3~$\upmu$m~$\times$~3~$\upmu$m to 10~$\upmu$m~$\times$~10~$\upmu$m, by electron-beam (e-beam) lithography. The heterostructures were etched by reactive-ion etching (RIE) employing an 80-nm-thick Al hard mask deposited in an e-beam evaporator. Graphene and hBN were etched selectively by O$_2$ and SF$_6$ plasma, respectively, to expose edges of graphene. The hard mask was removed in tetramethylammonium hydroxide solution after etching. 

The graphene edges were exposed along all four sides of the square-shaped heterostructures. Two of the graphene edges exposed on the opposite sides of the square were metallized to realize electrical contacts~\cite{wang13}. The Cr/Pd/Au (2/5/80~nm) contacts were patterned by e-beam lithography and slightly overlapped the heterostructures. They were deposited in an e-beam evaporator at the base pressure of $10^{-6}$~mbar. Fabricated devices were wire-bonded to chip-carriers for the electrical and PC measurements.

\section{Transport and PC measurements}\label{Section:Transport_and_PC_measurements}
\subsection{Transport measurements}
\begin{figure}
\centering
\includegraphics[scale=1.3]{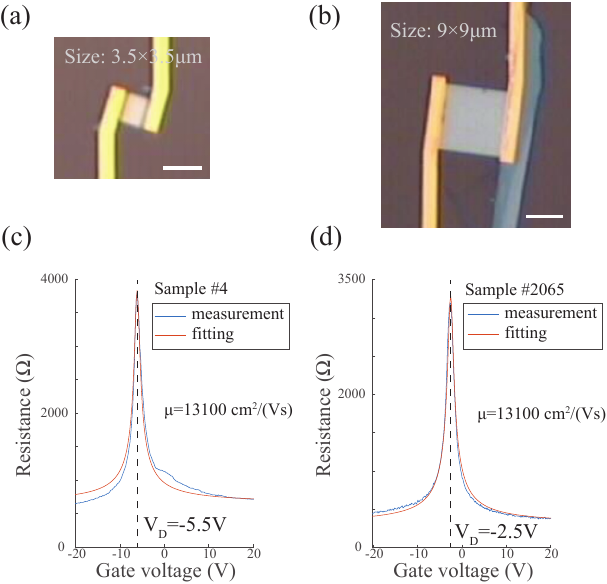}
\caption{Optical images of the two studied samples: sample \#4 and \#2065, are shown in (a) and (b). White scale bars: 5~$\upmu$m. Corresponding transport measurements of the two samples at 4~K are shown in (c) and (d) with fittings. Their Dirac points are located at $V_\mathrm{D}=-5.5, -2.5$ V respectively. Contact resistance for sample \#4 is $650~\Omega$ whereas $288~\Omega$ for sample \#2065.}
\label{SI_optical_images_and_mobillities}
\end{figure}

In Fig.~(\ref{SI_optical_images_and_mobillities}), we show optical pictures and mobility fittings with transport measurements of the two samples at 4~K. Two samples are of sizes $3.5\times 3.5~\upmu m$ and $9 \times 9~\upmu m$ respectively and the fitted mobilities of both samples~\cite{kim2009realization} are about $13000~\mathrm{cm^2/Vs}$.

Standard transport measurements are obtained with low frequency lock-in technique with an excitation current of 20~nA at 13~Hz. Each transport data point is an average of 10 measurements. Gate voltage sweeping is achieved using a DC source measure unit (Keithley 2400). The gate voltage is ramped with a step size of 5~mV and between two steps, a wait time of 10~ms is added to avoid hysteresis. Every 10 steps, we wait another 200~ms. The sample resistance is measured every 20~mV or 100~mV depending on the total scan range of gate voltage.

\begin{figure}
\centering
\includegraphics[scale=1.1]{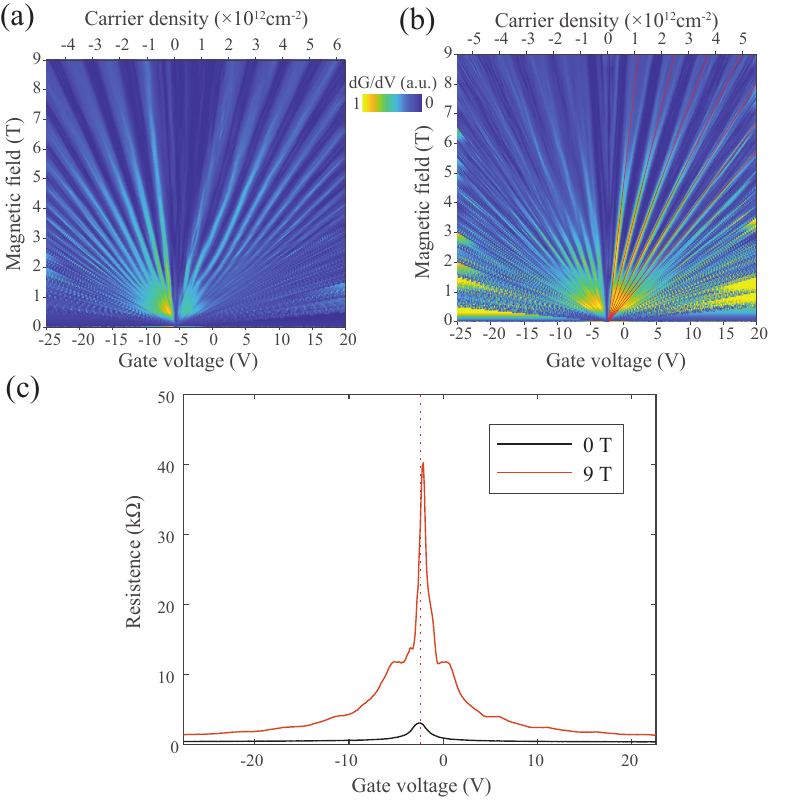}
\caption{Transport LL fan are shown for both samples: (a) for sample \#4; (b) for sample \#2065. We plot $dG/dV$ since the samples are two-terminal~\cite{bolotin2009observation}. The red lines are fitted even-fillings which are used in Fig.~(\ref{Fig2})(a). In (c), we plot the resistance for sample \#2065 measured at 9 T. 0 T result is included for comparison. In the transport measurements, there is no miniband appearing and the feature corresponding to $\mathrm{LL}_0$ does not split, indicating that there is no symmetry-broken states near the Dirac point. The Dirac point is marked by the vertical dashed line.}
\label{SI_Transport_LL_fan_two_samples}
\end{figure}

The Landau fans for the two samples, obtained from transport measurements within magnetic fields, are shown in Fig.~(\ref{SI_Transport_LL_fan_two_samples}). Since the samples are two-terminal, the measured is a combination of Hall and the longitudinal conductance~\cite{abanin2008conformal,williams2009quantum,sarma2011electronic} and we plot the derivative with respect to gate voltage $dG/dV$. In Fig.~(\ref{SI_Transport_LL_fan_two_samples})(b), we fit the even-fillings of the LLs and use them to benchmark the PC Landau fan plotted in Fig.~(\ref{Fig2})(a). In addition, we plot the transport measurement at 9 T in Fig.~(\ref{SI_Transport_LL_fan_two_samples})(c) which shows no symmetry-broken features. Transport measurements with/without optical illumination show no difference.

\subsection{TIA}\label{Subsection:TIA}
\begin{figure}
\centering
\includegraphics[scale=0.4]{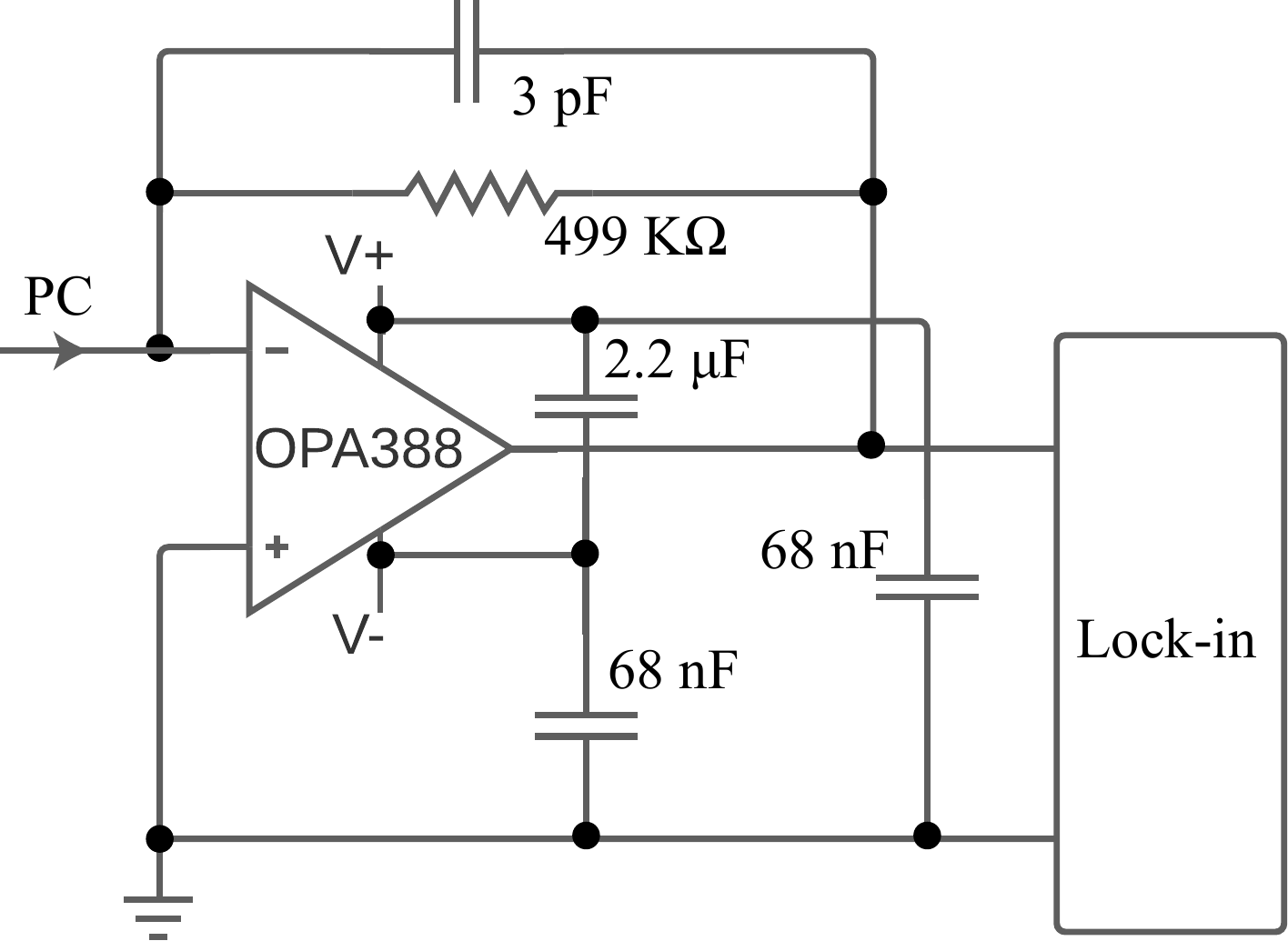}
\caption{Circuit design of the TIA. Three capacitors of 2.2 $\upmu$F, 68 nF and 68 nF are used to stabilize the circuit. The PC is measured using the TIA and the output voltage is measured with the voltage channels of a frequency-locked lock-in amplifier.}
\label{Design_of_the_TIA}
\end{figure}

In Fig.~(\ref{Design_of_the_TIA}), we show the circuit of the home-made TIA. Calibrations show that the TIA has a linear conversion from current (0 to 100~nA) to voltage with a constant ratio which is independent of the sample impedance.

\begin{figure}
\centering
\includegraphics[scale=0.8]{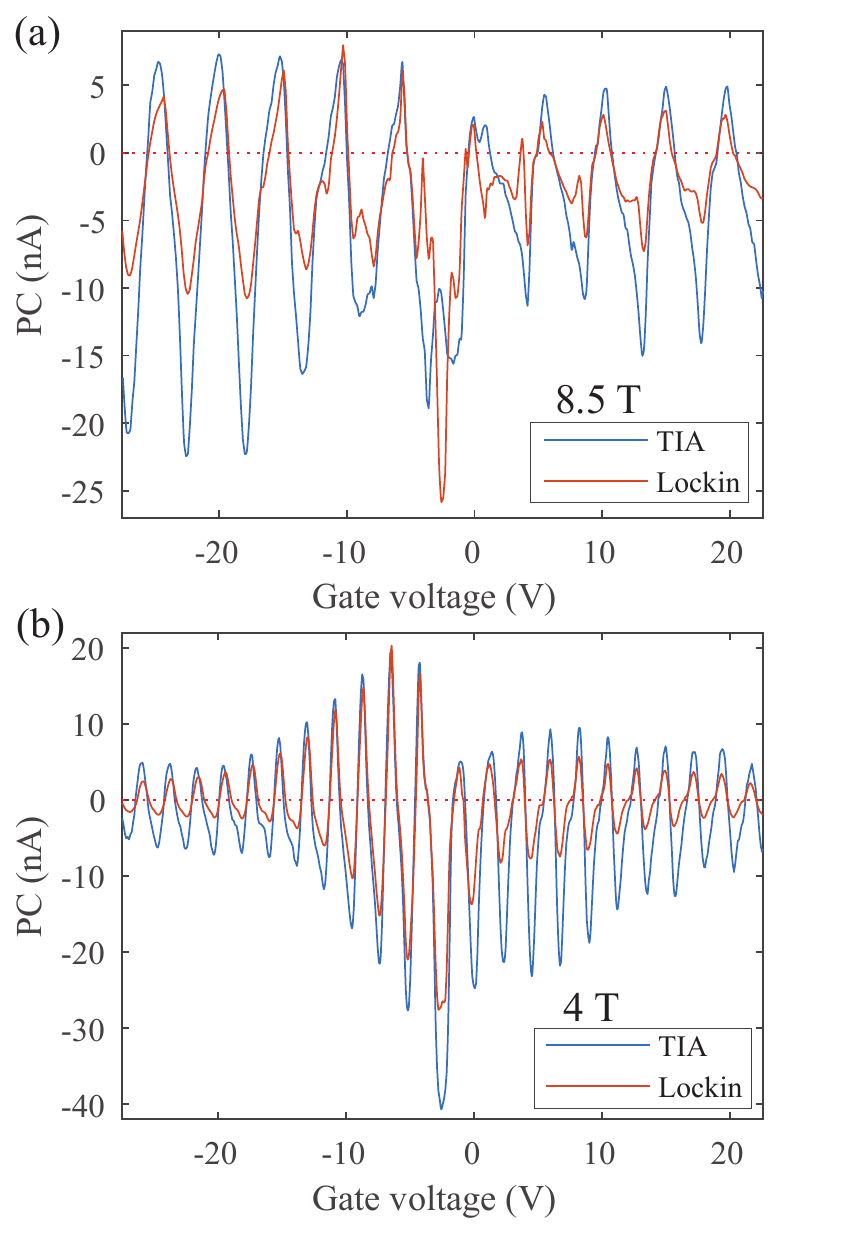}
\caption{To illustrate the importance of the TIA, here we show comparison between the PC measured through the TIA versus with the lock-in amplifier's current channel. In (a), we show PCs measured at 8.5 T for sample \#2065. In (b), we show PCs measured at 4 T for the same sample. Significant difference in the envelope is observed. In addition, PC measured without the TIA cannot recover the dip at the Dirac point at 8.5 T. These comparisons demonstrate that a TIA is necessary for precise PC measurements in graphene in the quantum Hall regime.}
\label{SI_Comparison_TIA_and_Lockin_4T_8T}
\end{figure}

The function of TIA is to remove the impact of the substantially varying sample resistance~\cite{horowitz2002art} from the measurements of PC. In the quantum Hall regime, the resistance of a two-terminal rectangular graphene sample is given by the quantized value $R_K/(4\nu+2)$ within the plateaus regime below $\mathrm{LL}_{\nu+1}$, where $R_K=h/e^2 \approx 26~\mathrm{k\Omega}$~\cite{zhang2005experimental,goerbig2011electronic} and the resistance is even higher near the Dirac point, as shown in Fig.~(\ref{SI_Transport_LL_fan_two_samples})(c). In comparison, the input impedance of common commercial current preamplifier (e.g. SR570) with a similar sensitivity is between 100~$\Omega$ and 10~k$\Omega$, which is comparable to the sample's impedance. Similarly, the input impedance of the current channel of lock-in amplifiers is also high: 1 k$\Omega$ for SR530,  100 $\Omega$/1 k$\Omega$ for SR860. As a result, when $E_\mathrm{F}$ is swept through many LLs, the impedance of the sample changes dramatically and thus an artificial envelope is imposed to the measured PC, making measurements inaccurate if without a TIA. Comparison of PC measurements with and without TIA is plotted in Fig.~(\ref{SI_Comparison_TIA_and_Lockin_4T_8T}), where the inaccuracy of measuring PC directly with a lock-in amplifier is clearly demonstrated. 

To show the function of TIA in the PC measurements, we compare PC obtained through a TIA and directly from the current channel of the lock-in amplifier (SR530), which has an impedance of 1~k$\Omega$. At a high magnetic field of 8.5 T, measurement without TIA, as shown in Fig.~(\ref{SI_Comparison_TIA_and_Lockin_4T_8T})(a), shows weaker PC values on the majority carrier side and the PC peaks are not asymmetric. This does not reflect the mechanism of carriers' relaxation as discussed in the main text. On the other hand, measurements using a TIA shows less noise. This is expected as a TIA does no exhibit the usual Johnson–Nyquist noise~\cite{donati2021photodetectors}. In addition, the PC dip at the Dirac point revealed by TIA measurements does not appear because of the divergence of sample's longitudinal impedance at the Dirac point~\cite{abanin2007dissipative,sarma2011electronic}, obscuring another important feature. The data at low field strength is shown in Fig.~(\ref{SI_Comparison_TIA_and_Lockin_4T_8T})(b). Measurements without a TIA shows a faster decaying envelope of PC oscillations, as $E_\mathrm{F}$ is moved away from the Dirac point, compared to the accurate one measured using a TIA.

\subsection{PC measurements}
PC measurements are obtained with a chopped laser at 308~Hz. The intensity, polarization, wavelength of the laser are stabilized at a frequency of 30~Hz to ensure accurate measurements. Further information are detailed below. The sample, inside a variable temperature insert (VTI), is mounted on top of a piezo-electric stack (scanners (ANSxy100) and positioners (ANPx101, ANPz201), with a total resolution of sub-nm). It is cooled down to 4~K and an out-of-plane magnetic field up to 9~T can be applied. The VTI has an optical window on top and a confocal microscope is built above to optically resolve the sample. The excitation laser is illuminated through the same window and the laser spot's position with respect to the sample can be monitored with the microscope and can be adjusted with the stack as well as a pizeo-controlled mirror mount. The sample is not biased. The two electrodes of the sample are connected to the TIA located outside the cryostat and the outputs of the TIA are connected with the frequncy-locked lock-in amplifier (SR860). Each data point of the PC is an average of 10 measurements. Gate sweeping is achieved the same way as the transport measurement described above.

\begin{figure}
\centering
\includegraphics[scale=0.5]{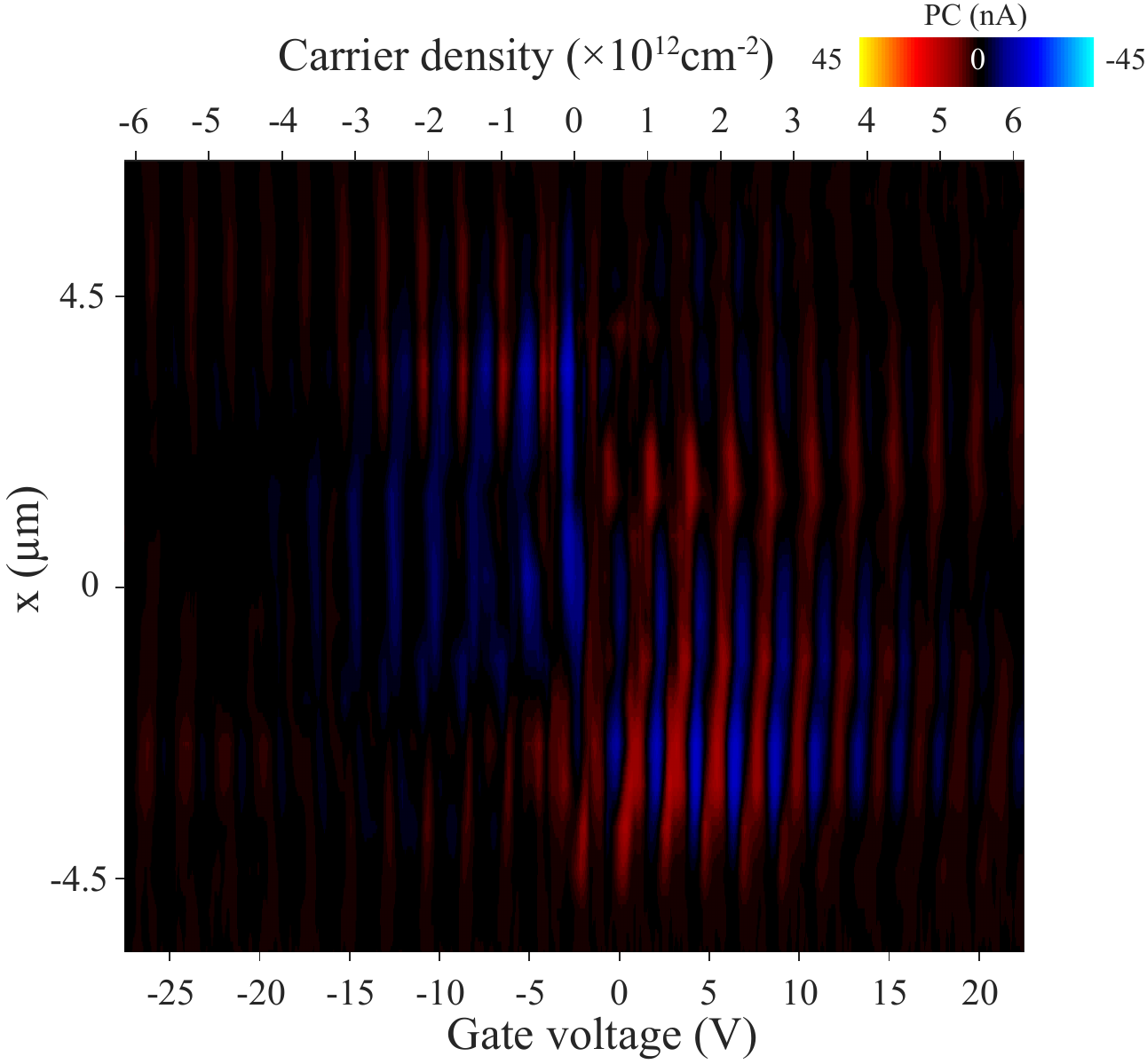}
\caption{The sum of the PCs measured at +4.5~T and -4.5~T, representing the $B$-independent part of PC, as a function of $x$ and gate voltage. The amplitude of the sum is minimal, compared to the subtraction shown in Fig.~(\ref{Fig1})(b).}
\label{SI_Chip4_sample2065_4T_sum}
\end{figure}

In order to obtain the $B$-dependent and independent parts of the PC, we subtract and sum the PC measured at +$B$ and -$B$: [PC(+$B$)-PC(-$B$)]/2; [PC(+$B$)-PC(-$B$)]/2. The substation has been shown in Fig.~(\ref{Fig1})(b). The sum is shown in Fig.~(\ref{SI_Chip4_sample2065_4T_sum}) representing the $B$-independent part of PC. As one can see, the $B$-independent part of PC is minimal compared to the dependent part, confirming that the transport of carriers are prominently through chiral edge states. The $B$-independent part of PC is attributed to the direct diffusion of carriers to the contacts~\cite{song2014shockley}.

\begin{figure}
\centering
\includegraphics[scale=1]{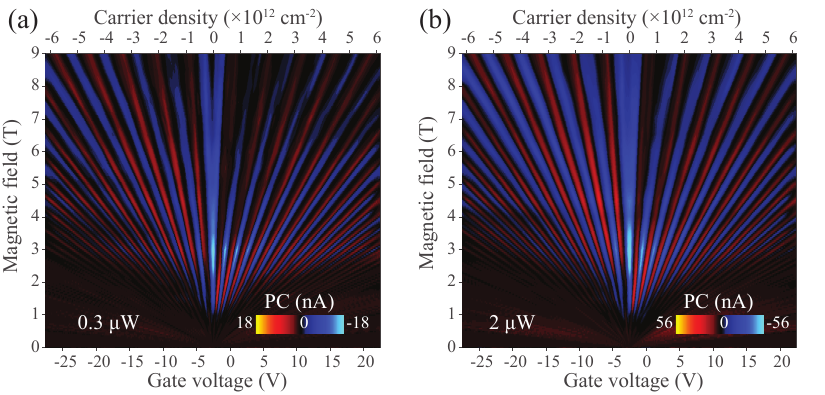}
\caption{PC Landau fan for sample \#2065 measured at 0.3 $\upmu$W and 2 $\upmu$W, shown as in (a) and (b) respectively, showing the same oscillation behavior as reported in the main text. }
\label{SI_PC_LL_fan_4_17_2020_chip4sample2065_0.3uW_and_2uW}
\end{figure}

The PC as a function of gate voltage and field strength (PC Landau fan) for sample \#2065, measured at a laser power of 1~$\upmu$W, has been shown in Fig.~(\ref{Fig2}). We measure the PC Landau fan with a laser power up to 8 $\upmu$W and the same oscillation behavior, as reported in the main text, is reliably observed. For example, PC Landau fan for laser powers of 0.3 $\upmu$W and 2 $\upmu$W are shown in Fig.~(\ref{SI_PC_LL_fan_4_17_2020_chip4sample2065_0.3uW_and_2uW}). In addition, we see minimal PC when $B=0$, distinguishing our observations from PC generated by inhomogeneity or PN junctions~\cite{gabor2011hot}. 

\begin{figure}
\centering
\includegraphics[scale=1]{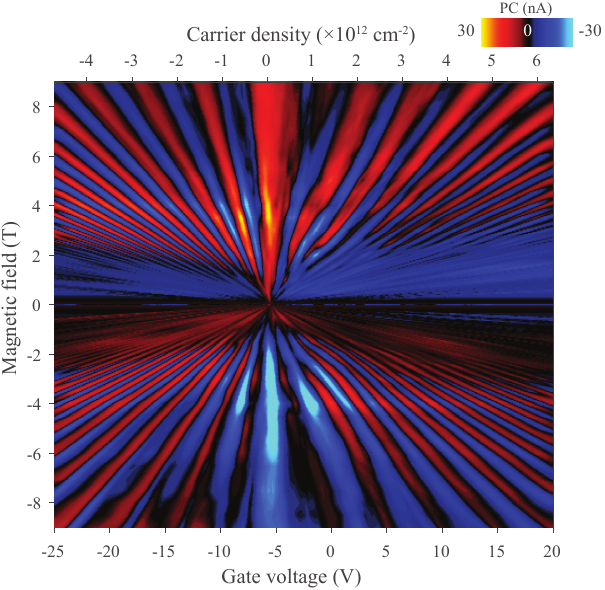}
\caption{PC Landau fan for sample \#4. Features agree with the presented analysis in the main text, except the missing dip at the Dirac point at high field strength. This is expected because this particular PC Landau fan was firstly measured directly using the lock-in and then corrected using the parameters of the TIA and thus the dip in the 0th LL at high field strength does not appear as a result of the obscuration with the sample impedance.}
\label{SI_Correted_PC_fan_chip1_sample4}
\end{figure}

We also show the PC Landau fan for sample \#4 in Fig.~(\ref{SI_Correted_PC_fan_chip1_sample4}). Features discussed in the main text are repeated, except the dip of the PC at the Dirac point. This is because Fig.~(\ref{SI_Correted_PC_fan_chip1_sample4}) is obtained directly with the lock-in amplifier's current channel without a TIA and calibrated afterwards to reflect the accurate envelope. However the dip at the Dirac point cannot be recovered by a simple calibration.

\begin{figure}
\centering
\includegraphics[scale=0.8]{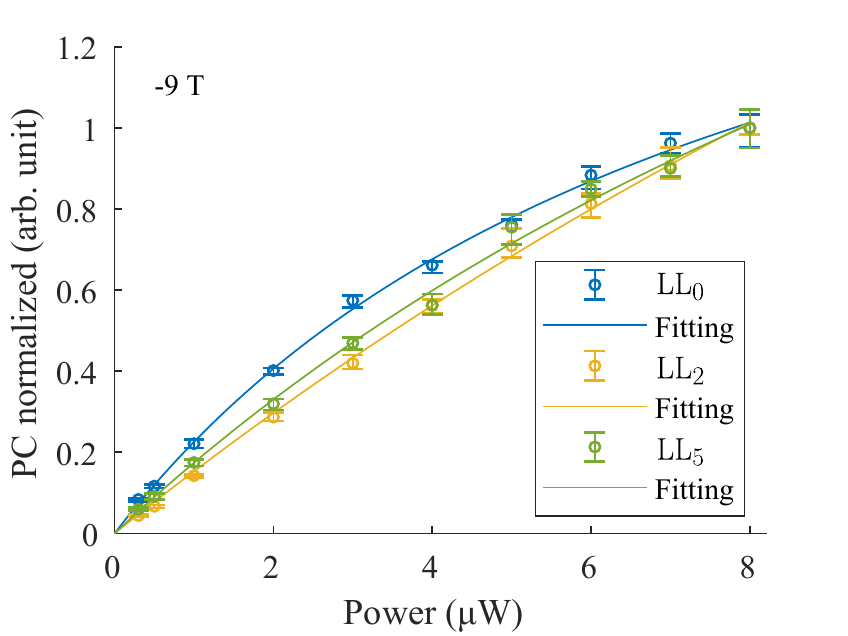}
\caption{We plot PC peaks' scaling with the laser power for various LLs, at a high magnetic field strength of -9 T. We observe that at high field strength, PC on the majority carrier side do not saturate, whereas saturations are observed for a low magnetic field as shown in Fig.~(\ref{Fig3})(a).}
\label{SI_Saturation_3_7_2020_chip1_sample4_-9T}
\end{figure}

To accompany the saturation behavior discussion, we show PC's scaling on laser power at a high field strength of -9 T in Fig.~(\ref{SI_Saturation_3_7_2020_chip1_sample4_-9T}). In Fig.~(\ref{Fig3}), we have shown the saturation of the PC peaks with laser power at a low field strength of -3.5 T, where saturations are observed for LLs with high index. In comparison, PC peaks at -9~T hardly saturate as shown in Fig.~(\ref{SI_Saturation_3_7_2020_chip1_sample4_-9T}), for those LLs within the reach of the gate.

\subsection{Laser power and polarization stabilization}
The laser power is constantly monitored with a photodiode right before the beam enters the optical window and a feedback is given to a PID loop controlling a laser power control module. The power control module is home-made using a liquid-crystal retarder and a polarizer. The frequency of the power stabilization is tuned such that the power is still chopped mechanically at 308~Hz while remains stable over each measurement duration, which is usually longer than 1~s. In total, power fluctuation can be controlled within 0.5\%. The polarization stabilization is achieved by diffracting the beam on a grating imprinted on a liquid-crystal spatial light modulator, which only works with a linear polarized light, and we only use the first order of diffraction. This ensures the polarization of the beam is clean and this is done before the photodiode used for the power stabilization. 

\subsection{Wavelength dependence PC measurements}

\begin{figure}[H]
\centering
\includegraphics[scale=0.8]{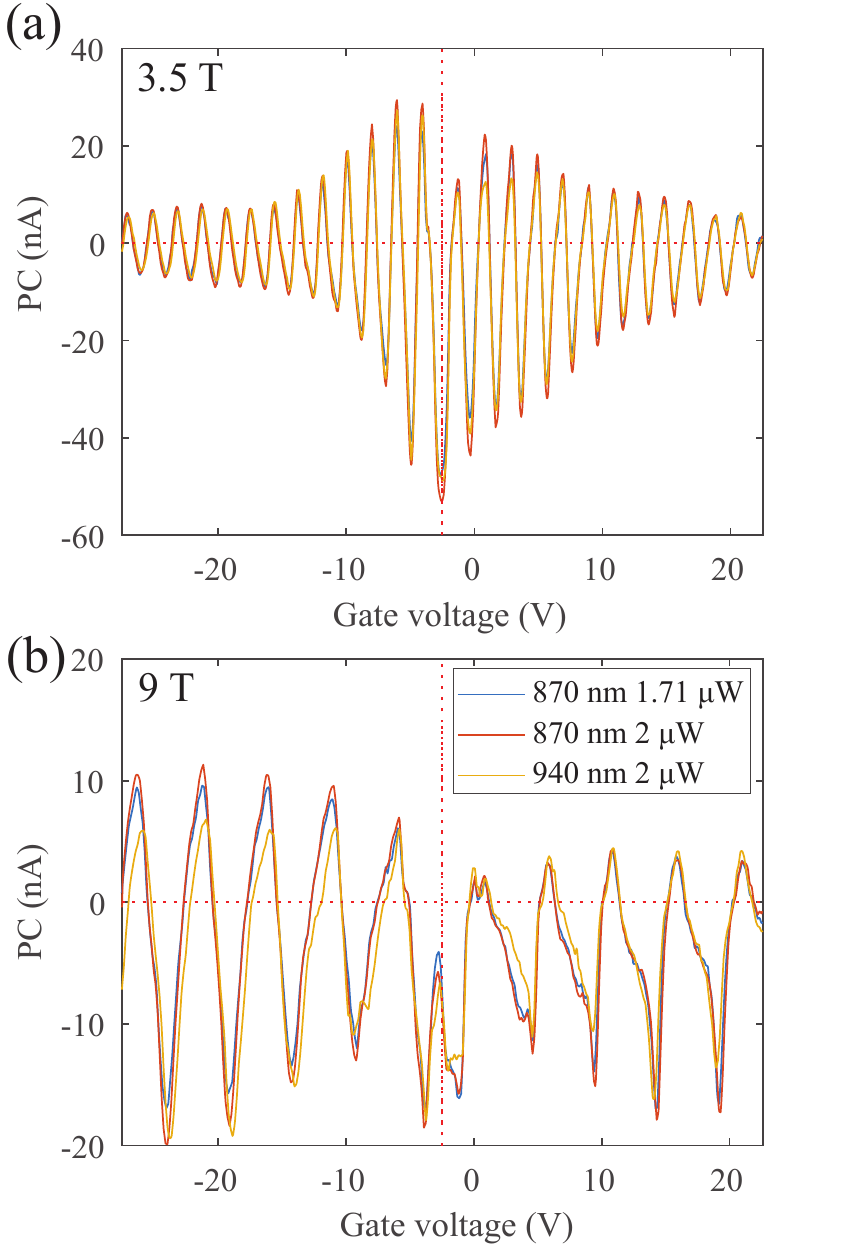}
\caption{PC measurements with excitation wavelength of 870~nm with laser powers 1.71, 2~$\upmu$W and 940~nm with 2~$\upmu$W are shown. In (a), we show measurements at 3.5~T whereas 9~T results are shown in (b). PCs measured at excitation wavelength of 870 nm and 940 nm overlap well, indicating that LLs, resonant to the excitation wavelength, merge together.}
\label{SI_Wavelength_dependence_chip4_sample2065_870_940nm_3d5T_and_9T}
\end{figure}

By changing excitation wavelength from 940~nm to 870~nm, we see little change in the PC measurements. In Fig.~(\ref{SI_Wavelength_dependence_chip4_sample2065_870_940nm_3d5T_and_9T}), we show measurements at 940~nm with a power of 2~$\upmu$W and at 870~nm with powers of 1.71 and 2~$\upmu$W. Measurements with different wavelengths match well, whereas PC peaks at different LLs show different scalings with power as explained in the main text.

The reason for selecting 1.71~$\upmu$W to compare with 2~$\upmu$W is the following. Since the wavelength is changed, the laser focal spot size is changed accordingly. To maintain the average intensity same as 940~nm with 2~$\upmu$W, one power for 870~nm is chosen as $(870/940)^2\times 2~\upmu\mathrm{W}=1.71~\upmu$W.

\bibliography{PC_measurements}

\begin{thebibliography}{72}%
\makeatletter
\providecommand \@ifxundefined [1]{%
 \@ifx{#1\undefined}
}%
\providecommand \@ifnum [1]{%
 \ifnum #1\expandafter \@firstoftwo
 \else \expandafter \@secondoftwo
 \fi
}%
\providecommand \@ifx [1]{%
 \ifx #1\expandafter \@firstoftwo
 \else \expandafter \@secondoftwo
 \fi
}%
\providecommand \natexlab [1]{#1}%
\providecommand \enquote  [1]{``#1''}%
\providecommand \bibnamefont  [1]{#1}%
\providecommand \bibfnamefont [1]{#1}%
\providecommand \citenamefont [1]{#1}%
\providecommand \href@noop [0]{\@secondoftwo}%
\providecommand \href [0]{\begingroup \@sanitize@url \@href}%
\providecommand \@href[1]{\@@startlink{#1}\@@href}%
\providecommand \@@href[1]{\endgroup#1\@@endlink}%
\providecommand \@sanitize@url [0]{\catcode `\\12\catcode `\$12\catcode
  `\&12\catcode `\#12\catcode `\^12\catcode `\_12\catcode `\%12\relax}%
\providecommand \@@startlink[1]{}%
\providecommand \@@endlink[0]{}%
\providecommand \url  [0]{\begingroup\@sanitize@url \@url }%
\providecommand \@url [1]{\endgroup\@href {#1}{\urlprefix }}%
\providecommand \urlprefix  [0]{URL }%
\providecommand \Eprint [0]{\href }%
\providecommand \doibase [0]{http://dx.doi.org/}%
\providecommand \selectlanguage [0]{\@gobble}%
\providecommand \bibinfo  [0]{\@secondoftwo}%
\providecommand \bibfield  [0]{\@secondoftwo}%
\providecommand \translation [1]{[#1]}%
\providecommand \BibitemOpen [0]{}%
\providecommand \bibitemStop [0]{}%
\providecommand \bibitemNoStop [0]{.\EOS\space}%
\providecommand \EOS [0]{\spacefactor3000\relax}%
\providecommand \BibitemShut  [1]{\csname bibitem#1\endcsname}%
\let\auto@bib@innerbib\@empty
%</preamble>
\bibitem [{\citenamefont {Hasan}\ and\ \citenamefont
  {Kane}(2010)}]{hasan2010colloquium}%
  \BibitemOpen
  \bibfield  {author} {\bibinfo {author} {\bibfnamefont {M.~Z.}\ \bibnamefont
  {Hasan}}\ and\ \bibinfo {author} {\bibfnamefont {C.~L.}\ \bibnamefont
  {Kane}},\ }\href@noop {} {\bibfield  {journal} {\bibinfo  {journal} {Reviews
  of modern physics}\ }\textbf {\bibinfo {volume} {82}},\ \bibinfo {pages}
  {3045} (\bibinfo {year} {2010})}\BibitemShut {NoStop}%
\bibitem [{\citenamefont {Qi}\ and\ \citenamefont
  {Zhang}(2011)}]{qi2011topological}%
  \BibitemOpen
  \bibfield  {author} {\bibinfo {author} {\bibfnamefont {X.-L.}\ \bibnamefont
  {Qi}}\ and\ \bibinfo {author} {\bibfnamefont {S.-C.}\ \bibnamefont {Zhang}},\
  }\href@noop {} {\bibfield  {journal} {\bibinfo  {journal} {Reviews of Modern
  Physics}\ }\textbf {\bibinfo {volume} {83}},\ \bibinfo {pages} {1057}
  (\bibinfo {year} {2011})}\BibitemShut {NoStop}%
\bibitem [{\citenamefont {Halperin}(1982)}]{halperin1982quantized}%
  \BibitemOpen
  \bibfield  {author} {\bibinfo {author} {\bibfnamefont {B.~I.}\ \bibnamefont
  {Halperin}},\ }\href@noop {} {\bibfield  {journal} {\bibinfo  {journal}
  {Physical Review B}\ }\textbf {\bibinfo {volume} {25}},\ \bibinfo {pages}
  {2185} (\bibinfo {year} {1982})}\BibitemShut {NoStop}%
\bibitem [{\citenamefont {Jackiw}\ and\ \citenamefont
  {Rebbi}(1976)}]{jackiw1976solitons}%
  \BibitemOpen
  \bibfield  {author} {\bibinfo {author} {\bibfnamefont {R.}~\bibnamefont
  {Jackiw}}\ and\ \bibinfo {author} {\bibfnamefont {C.}~\bibnamefont {Rebbi}},\
  }\href@noop {} {\bibfield  {journal} {\bibinfo  {journal} {Physical Review
  D}\ }\textbf {\bibinfo {volume} {13}},\ \bibinfo {pages} {3398} (\bibinfo
  {year} {1976})}\BibitemShut {NoStop}%
\bibitem [{\citenamefont {Girvin}(1999)}]{girvin1999quantum}%
  \BibitemOpen
  \bibfield  {author} {\bibinfo {author} {\bibfnamefont {S.~M.}\ \bibnamefont
  {Girvin}},\ }in\ \href@noop {} {\emph {\bibinfo {booktitle} {Aspects
  topologiques de la physique en basse dimension. Topological aspects of low
  dimensional systems}}}\ (\bibinfo  {publisher} {Springer},\ \bibinfo {year}
  {1999})\ pp.\ \bibinfo {pages} {53--175}\BibitemShut {NoStop}%
\bibitem [{\citenamefont {Patlatiuk}\ \emph {et~al.}(2018)\citenamefont
  {Patlatiuk}, \citenamefont {Scheller}, \citenamefont {Hill}, \citenamefont
  {Tserkovnyak}, \citenamefont {Barak}, \citenamefont {Yacoby}, \citenamefont
  {Pfeiffer}, \citenamefont {West},\ and\ \citenamefont
  {Zumb{\"u}hl}}]{patlatiuk2018evolution}%
  \BibitemOpen
  \bibfield  {author} {\bibinfo {author} {\bibfnamefont {T.}~\bibnamefont
  {Patlatiuk}}, \bibinfo {author} {\bibfnamefont {C.~P.}\ \bibnamefont
  {Scheller}}, \bibinfo {author} {\bibfnamefont {D.}~\bibnamefont {Hill}},
  \bibinfo {author} {\bibfnamefont {Y.}~\bibnamefont {Tserkovnyak}}, \bibinfo
  {author} {\bibfnamefont {G.}~\bibnamefont {Barak}}, \bibinfo {author}
  {\bibfnamefont {A.}~\bibnamefont {Yacoby}}, \bibinfo {author} {\bibfnamefont
  {L.~N.}\ \bibnamefont {Pfeiffer}}, \bibinfo {author} {\bibfnamefont {K.~W.}\
  \bibnamefont {West}}, \ and\ \bibinfo {author} {\bibfnamefont {D.~M.}\
  \bibnamefont {Zumb{\"u}hl}},\ }\href@noop {} {\bibfield  {journal} {\bibinfo
  {journal} {Nature communications}\ }\textbf {\bibinfo {volume} {9}},\
  \bibinfo {pages} {1} (\bibinfo {year} {2018})}\BibitemShut {NoStop}%
\bibitem [{\citenamefont {Williams}\ \emph {et~al.}(2007)\citenamefont
  {Williams}, \citenamefont {DiCarlo},\ and\ \citenamefont
  {Marcus}}]{williams2007quantum}%
  \BibitemOpen
  \bibfield  {author} {\bibinfo {author} {\bibfnamefont {J.}~\bibnamefont
  {Williams}}, \bibinfo {author} {\bibfnamefont {L.}~\bibnamefont {DiCarlo}}, \
  and\ \bibinfo {author} {\bibfnamefont {C.}~\bibnamefont {Marcus}},\
  }\href@noop {} {\bibfield  {journal} {\bibinfo  {journal} {Science}\ }\textbf
  {\bibinfo {volume} {317}},\ \bibinfo {pages} {638} (\bibinfo {year}
  {2007})}\BibitemShut {NoStop}%
\bibitem [{\citenamefont {Abanin}\ and\ \citenamefont
  {Levitov}(2007)}]{abanin2007quantized}%
  \BibitemOpen
  \bibfield  {author} {\bibinfo {author} {\bibfnamefont {D.}~\bibnamefont
  {Abanin}}\ and\ \bibinfo {author} {\bibfnamefont {L.}~\bibnamefont
  {Levitov}},\ }\href@noop {} {\bibfield  {journal} {\bibinfo  {journal}
  {Science}\ }\textbf {\bibinfo {volume} {317}},\ \bibinfo {pages} {641}
  (\bibinfo {year} {2007})}\BibitemShut {NoStop}%
\bibitem [{\citenamefont {Nazin}\ \emph {et~al.}(2010)\citenamefont {Nazin},
  \citenamefont {Zhang}, \citenamefont {Zhang}, \citenamefont {Sutter},\ and\
  \citenamefont {Sutter}}]{nazin2010visualization}%
  \BibitemOpen
  \bibfield  {author} {\bibinfo {author} {\bibfnamefont {G.}~\bibnamefont
  {Nazin}}, \bibinfo {author} {\bibfnamefont {Y.}~\bibnamefont {Zhang}},
  \bibinfo {author} {\bibfnamefont {L.}~\bibnamefont {Zhang}}, \bibinfo
  {author} {\bibfnamefont {E.}~\bibnamefont {Sutter}}, \ and\ \bibinfo {author}
  {\bibfnamefont {P.}~\bibnamefont {Sutter}},\ }\href@noop {} {\bibfield
  {journal} {\bibinfo  {journal} {Nature Physics}\ }\textbf {\bibinfo {volume}
  {6}},\ \bibinfo {pages} {870} (\bibinfo {year} {2010})}\BibitemShut {NoStop}%
\bibitem [{\citenamefont {Queisser}\ and\ \citenamefont
  {Sch{\"u}tzhold}(2013)}]{queisser2013strong}%
  \BibitemOpen
  \bibfield  {author} {\bibinfo {author} {\bibfnamefont {F.}~\bibnamefont
  {Queisser}}\ and\ \bibinfo {author} {\bibfnamefont {R.}~\bibnamefont
  {Sch{\"u}tzhold}},\ }\href@noop {} {\bibfield  {journal} {\bibinfo  {journal}
  {Physical review letters}\ }\textbf {\bibinfo {volume} {111}},\ \bibinfo
  {pages} {046601} (\bibinfo {year} {2013})}\BibitemShut {NoStop}%
\bibitem [{\citenamefont {Abanin}\ \emph
  {et~al.}(2007{\natexlab{a}})\citenamefont {Abanin}, \citenamefont {Lee},\
  and\ \citenamefont {Levitov}}]{abanin2007charge}%
  \BibitemOpen
  \bibfield  {author} {\bibinfo {author} {\bibfnamefont {D.~A.}\ \bibnamefont
  {Abanin}}, \bibinfo {author} {\bibfnamefont {P.~A.}\ \bibnamefont {Lee}}, \
  and\ \bibinfo {author} {\bibfnamefont {L.~S.}\ \bibnamefont {Levitov}},\
  }\href@noop {} {\bibfield  {journal} {\bibinfo  {journal} {Solid state
  communications}\ }\textbf {\bibinfo {volume} {143}},\ \bibinfo {pages} {77}
  (\bibinfo {year} {2007}{\natexlab{a}})}\BibitemShut {NoStop}%
\bibitem [{\citenamefont {Abanin}\ \emph
  {et~al.}(2007{\natexlab{b}})\citenamefont {Abanin}, \citenamefont
  {Novoselov}, \citenamefont {Zeitler}, \citenamefont {Lee}, \citenamefont
  {Geim},\ and\ \citenamefont {Levitov}}]{abanin2007dissipative}%
  \BibitemOpen
  \bibfield  {author} {\bibinfo {author} {\bibfnamefont {D.~A.}\ \bibnamefont
  {Abanin}}, \bibinfo {author} {\bibfnamefont {K.~S.}\ \bibnamefont
  {Novoselov}}, \bibinfo {author} {\bibfnamefont {U.}~\bibnamefont {Zeitler}},
  \bibinfo {author} {\bibfnamefont {P.~A.}\ \bibnamefont {Lee}}, \bibinfo
  {author} {\bibfnamefont {A.~K.}\ \bibnamefont {Geim}}, \ and\ \bibinfo
  {author} {\bibfnamefont {L.~S.}\ \bibnamefont {Levitov}},\ }\href@noop {}
  {\bibfield  {journal} {\bibinfo  {journal} {Physical review letters}\
  }\textbf {\bibinfo {volume} {98}},\ \bibinfo {pages} {196806} (\bibinfo
  {year} {2007}{\natexlab{b}})}\BibitemShut {NoStop}%
\bibitem [{\citenamefont {Kim}\ \emph {et~al.}(2021)\citenamefont {Kim},
  \citenamefont {Schwenk}, \citenamefont {Walkup}, \citenamefont {Zeng},
  \citenamefont {Ghahari}, \citenamefont {Le}, \citenamefont {Slot},
  \citenamefont {Berwanger}, \citenamefont {Blankenship}, \citenamefont
  {Watanabe} \emph {et~al.}}]{kim2021edge}%
  \BibitemOpen
  \bibfield  {author} {\bibinfo {author} {\bibfnamefont {S.}~\bibnamefont
  {Kim}}, \bibinfo {author} {\bibfnamefont {J.}~\bibnamefont {Schwenk}},
  \bibinfo {author} {\bibfnamefont {D.}~\bibnamefont {Walkup}}, \bibinfo
  {author} {\bibfnamefont {Y.}~\bibnamefont {Zeng}}, \bibinfo {author}
  {\bibfnamefont {F.}~\bibnamefont {Ghahari}}, \bibinfo {author} {\bibfnamefont
  {S.~T.}\ \bibnamefont {Le}}, \bibinfo {author} {\bibfnamefont {M.~R.}\
  \bibnamefont {Slot}}, \bibinfo {author} {\bibfnamefont {J.}~\bibnamefont
  {Berwanger}}, \bibinfo {author} {\bibfnamefont {S.~R.}\ \bibnamefont
  {Blankenship}}, \bibinfo {author} {\bibfnamefont {K.}~\bibnamefont
  {Watanabe}},  \emph {et~al.},\ }\href@noop {} {\bibfield  {journal} {\bibinfo
   {journal} {Nature Communications}\ }\textbf {\bibinfo {volume} {12}},\
  \bibinfo {pages} {1} (\bibinfo {year} {2021})}\BibitemShut {NoStop}%
\bibitem [{\citenamefont {Sonntag}\ \emph {et~al.}(2017)\citenamefont
  {Sonntag}, \citenamefont {Kurzmann}, \citenamefont {Geller}, \citenamefont
  {Queisser}, \citenamefont {Lorke},\ and\ \citenamefont
  {Sch{\"u}tzhold}}]{sonntag2017giant}%
  \BibitemOpen
  \bibfield  {author} {\bibinfo {author} {\bibfnamefont {J.}~\bibnamefont
  {Sonntag}}, \bibinfo {author} {\bibfnamefont {A.}~\bibnamefont {Kurzmann}},
  \bibinfo {author} {\bibfnamefont {M.}~\bibnamefont {Geller}}, \bibinfo
  {author} {\bibfnamefont {F.}~\bibnamefont {Queisser}}, \bibinfo {author}
  {\bibfnamefont {A.}~\bibnamefont {Lorke}}, \ and\ \bibinfo {author}
  {\bibfnamefont {R.}~\bibnamefont {Sch{\"u}tzhold}},\ }\href@noop {}
  {\bibfield  {journal} {\bibinfo  {journal} {New Journal of Physics}\ }\textbf
  {\bibinfo {volume} {19}},\ \bibinfo {pages} {063028} (\bibinfo {year}
  {2017})}\BibitemShut {NoStop}%
\bibitem [{\citenamefont {Cao}\ \emph {et~al.}(2016)\citenamefont {Cao},
  \citenamefont {Aivazian}, \citenamefont {Fei}, \citenamefont {Ross},
  \citenamefont {Cobden},\ and\ \citenamefont {Xu}}]{cao2016photo}%
  \BibitemOpen
  \bibfield  {author} {\bibinfo {author} {\bibfnamefont {H.}~\bibnamefont
  {Cao}}, \bibinfo {author} {\bibfnamefont {G.}~\bibnamefont {Aivazian}},
  \bibinfo {author} {\bibfnamefont {Z.}~\bibnamefont {Fei}}, \bibinfo {author}
  {\bibfnamefont {J.}~\bibnamefont {Ross}}, \bibinfo {author} {\bibfnamefont
  {D.~H.}\ \bibnamefont {Cobden}}, \ and\ \bibinfo {author} {\bibfnamefont
  {X.}~\bibnamefont {Xu}},\ }\href@noop {} {\bibfield  {journal} {\bibinfo
  {journal} {Nature Physics}\ }\textbf {\bibinfo {volume} {12}},\ \bibinfo
  {pages} {236} (\bibinfo {year} {2016})}\BibitemShut {NoStop}%
\bibitem [{\citenamefont {Wu}\ \emph {et~al.}(2016)\citenamefont {Wu},
  \citenamefont {Wang}, \citenamefont {Lai}, \citenamefont {Shan},
  \citenamefont {Aivazian}, \citenamefont {Zhang}, \citenamefont {Taniguchi},
  \citenamefont {Watanabe}, \citenamefont {Xiao}, \citenamefont {Dean} \emph
  {et~al.}}]{wu2016multiple}%
  \BibitemOpen
  \bibfield  {author} {\bibinfo {author} {\bibfnamefont {S.}~\bibnamefont
  {Wu}}, \bibinfo {author} {\bibfnamefont {L.}~\bibnamefont {Wang}}, \bibinfo
  {author} {\bibfnamefont {Y.}~\bibnamefont {Lai}}, \bibinfo {author}
  {\bibfnamefont {W.-Y.}\ \bibnamefont {Shan}}, \bibinfo {author}
  {\bibfnamefont {G.}~\bibnamefont {Aivazian}}, \bibinfo {author}
  {\bibfnamefont {X.}~\bibnamefont {Zhang}}, \bibinfo {author} {\bibfnamefont
  {T.}~\bibnamefont {Taniguchi}}, \bibinfo {author} {\bibfnamefont
  {K.}~\bibnamefont {Watanabe}}, \bibinfo {author} {\bibfnamefont
  {D.}~\bibnamefont {Xiao}}, \bibinfo {author} {\bibfnamefont {C.}~\bibnamefont
  {Dean}},  \emph {et~al.},\ }\href@noop {} {\bibfield  {journal} {\bibinfo
  {journal} {Science advances}\ }\textbf {\bibinfo {volume} {2}},\ \bibinfo
  {pages} {e1600002} (\bibinfo {year} {2016})}\BibitemShut {NoStop}%
\bibitem [{\citenamefont {Gazzano}\ \emph {et~al.}(2019)\citenamefont
  {Gazzano}, \citenamefont {Cao}, \citenamefont {Hu}, \citenamefont {Huber},
  \citenamefont {Grass}, \citenamefont {Gullans}, \citenamefont {Newell},
  \citenamefont {Hafezi},\ and\ \citenamefont
  {Solomon}}]{gazzano2019observation}%
  \BibitemOpen
  \bibfield  {author} {\bibinfo {author} {\bibfnamefont {O.}~\bibnamefont
  {Gazzano}}, \bibinfo {author} {\bibfnamefont {B.}~\bibnamefont {Cao}},
  \bibinfo {author} {\bibfnamefont {J.}~\bibnamefont {Hu}}, \bibinfo {author}
  {\bibfnamefont {T.}~\bibnamefont {Huber}}, \bibinfo {author} {\bibfnamefont
  {T.}~\bibnamefont {Grass}}, \bibinfo {author} {\bibfnamefont
  {M.}~\bibnamefont {Gullans}}, \bibinfo {author} {\bibfnamefont
  {D.}~\bibnamefont {Newell}}, \bibinfo {author} {\bibfnamefont
  {M.}~\bibnamefont {Hafezi}}, \ and\ \bibinfo {author} {\bibfnamefont {G.~S.}\
  \bibnamefont {Solomon}},\ }\href@noop {} {\bibfield  {journal} {\bibinfo
  {journal} {arXiv preprint arXiv:1903.01487}\ } (\bibinfo {year}
  {2019})}\BibitemShut {NoStop}%
\bibitem [{\citenamefont {Wu}(2016)}]{wu2016device}%
  \BibitemOpen
  \bibfield  {author} {\bibinfo {author} {\bibfnamefont {S.}~\bibnamefont
  {Wu}},\ }\emph {\bibinfo {title} {Device Physics of Two-Dimensional
  Crystalline Materials}},\ \href@noop {} {Ph.D. thesis} (\bibinfo {year}
  {2016})\BibitemShut {NoStop}%
\bibitem [{\citenamefont {Wendler}\ \emph {et~al.}(2014)\citenamefont
  {Wendler}, \citenamefont {Knorr},\ and\ \citenamefont
  {Malic}}]{wendler2014carrier}%
  \BibitemOpen
  \bibfield  {author} {\bibinfo {author} {\bibfnamefont {F.}~\bibnamefont
  {Wendler}}, \bibinfo {author} {\bibfnamefont {A.}~\bibnamefont {Knorr}}, \
  and\ \bibinfo {author} {\bibfnamefont {E.}~\bibnamefont {Malic}},\
  }\href@noop {} {\bibfield  {journal} {\bibinfo  {journal} {Nature
  communications}\ }\textbf {\bibinfo {volume} {5}},\ \bibinfo {pages} {3703}
  (\bibinfo {year} {2014})}\BibitemShut {NoStop}%
\bibitem [{\citenamefont {Wendler}\ \emph {et~al.}(2015)\citenamefont
  {Wendler}, \citenamefont {Knorr},\ and\ \citenamefont
  {Malic}}]{wendler2015ultrafast}%
  \BibitemOpen
  \bibfield  {author} {\bibinfo {author} {\bibfnamefont {F.}~\bibnamefont
  {Wendler}}, \bibinfo {author} {\bibfnamefont {A.}~\bibnamefont {Knorr}}, \
  and\ \bibinfo {author} {\bibfnamefont {E.}~\bibnamefont {Malic}},\
  }\href@noop {} {\bibfield  {journal} {\bibinfo  {journal} {Nanophotonics}\
  }\textbf {\bibinfo {volume} {4}},\ \bibinfo {pages} {224} (\bibinfo {year}
  {2015})}\BibitemShut {NoStop}%
\bibitem [{\citenamefont {Song}\ \emph {et~al.}(2013)\citenamefont {Song},
  \citenamefont {Tielrooij}, \citenamefont {Koppens},\ and\ \citenamefont
  {Levitov}}]{song2013photoexcited}%
  \BibitemOpen
  \bibfield  {author} {\bibinfo {author} {\bibfnamefont {J.~C.}\ \bibnamefont
  {Song}}, \bibinfo {author} {\bibfnamefont {K.~J.}\ \bibnamefont {Tielrooij}},
  \bibinfo {author} {\bibfnamefont {F.~H.}\ \bibnamefont {Koppens}}, \ and\
  \bibinfo {author} {\bibfnamefont {L.~S.}\ \bibnamefont {Levitov}},\
  }\href@noop {} {\bibfield  {journal} {\bibinfo  {journal} {Physical Review
  B}\ }\textbf {\bibinfo {volume} {87}},\ \bibinfo {pages} {155429} (\bibinfo
  {year} {2013})}\BibitemShut {NoStop}%
\bibitem [{\citenamefont {Tielrooij}\ \emph {et~al.}(2013)\citenamefont
  {Tielrooij}, \citenamefont {Song}, \citenamefont {Jensen}, \citenamefont
  {Centeno}, \citenamefont {Pesquera}, \citenamefont {Elorza}, \citenamefont
  {Bonn}, \citenamefont {Levitov},\ and\ \citenamefont
  {Koppens}}]{tielrooij2013photoexcitation}%
  \BibitemOpen
  \bibfield  {author} {\bibinfo {author} {\bibfnamefont {K.}~\bibnamefont
  {Tielrooij}}, \bibinfo {author} {\bibfnamefont {J.}~\bibnamefont {Song}},
  \bibinfo {author} {\bibfnamefont {S.~A.}\ \bibnamefont {Jensen}}, \bibinfo
  {author} {\bibfnamefont {A.}~\bibnamefont {Centeno}}, \bibinfo {author}
  {\bibfnamefont {A.}~\bibnamefont {Pesquera}}, \bibinfo {author}
  {\bibfnamefont {A.~Z.}\ \bibnamefont {Elorza}}, \bibinfo {author}
  {\bibfnamefont {M.}~\bibnamefont {Bonn}}, \bibinfo {author} {\bibfnamefont
  {L.}~\bibnamefont {Levitov}}, \ and\ \bibinfo {author} {\bibfnamefont
  {F.}~\bibnamefont {Koppens}},\ }\href@noop {} {\bibfield  {journal} {\bibinfo
   {journal} {Nature Physics}\ }\textbf {\bibinfo {volume} {9}},\ \bibinfo
  {pages} {248} (\bibinfo {year} {2013})}\BibitemShut {NoStop}%
\bibitem [{\citenamefont {Mittendorff}\ \emph {et~al.}(2015)\citenamefont
  {Mittendorff}, \citenamefont {Wendler}, \citenamefont {Malic}, \citenamefont
  {Knorr}, \citenamefont {Orlita}, \citenamefont {Potemski}, \citenamefont
  {Berger}, \citenamefont {De~Heer}, \citenamefont {Schneider}, \citenamefont
  {Helm} \emph {et~al.}}]{mittendorff2015carrier}%
  \BibitemOpen
  \bibfield  {author} {\bibinfo {author} {\bibfnamefont {M.}~\bibnamefont
  {Mittendorff}}, \bibinfo {author} {\bibfnamefont {F.}~\bibnamefont
  {Wendler}}, \bibinfo {author} {\bibfnamefont {E.}~\bibnamefont {Malic}},
  \bibinfo {author} {\bibfnamefont {A.}~\bibnamefont {Knorr}}, \bibinfo
  {author} {\bibfnamefont {M.}~\bibnamefont {Orlita}}, \bibinfo {author}
  {\bibfnamefont {M.}~\bibnamefont {Potemski}}, \bibinfo {author}
  {\bibfnamefont {C.}~\bibnamefont {Berger}}, \bibinfo {author} {\bibfnamefont
  {W.~A.}\ \bibnamefont {De~Heer}}, \bibinfo {author} {\bibfnamefont
  {H.}~\bibnamefont {Schneider}}, \bibinfo {author} {\bibfnamefont
  {M.}~\bibnamefont {Helm}},  \emph {et~al.},\ }\href@noop {} {\bibfield
  {journal} {\bibinfo  {journal} {Nature Physics}\ }\textbf {\bibinfo {volume}
  {11}},\ \bibinfo {pages} {75} (\bibinfo {year} {2015})}\BibitemShut {NoStop}%
\bibitem [{\citenamefont {Plotzing}\ \emph {et~al.}(2014)\citenamefont
  {Plotzing}, \citenamefont {Winzer}, \citenamefont {Malic}, \citenamefont
  {Neumaier}, \citenamefont {Knorr},\ and\ \citenamefont
  {Kurz}}]{plotzing2014experimental}%
  \BibitemOpen
  \bibfield  {author} {\bibinfo {author} {\bibfnamefont {T.}~\bibnamefont
  {Plotzing}}, \bibinfo {author} {\bibfnamefont {T.}~\bibnamefont {Winzer}},
  \bibinfo {author} {\bibfnamefont {E.}~\bibnamefont {Malic}}, \bibinfo
  {author} {\bibfnamefont {D.}~\bibnamefont {Neumaier}}, \bibinfo {author}
  {\bibfnamefont {A.}~\bibnamefont {Knorr}}, \ and\ \bibinfo {author}
  {\bibfnamefont {H.}~\bibnamefont {Kurz}},\ }\href@noop {} {\bibfield
  {journal} {\bibinfo  {journal} {Nano letters}\ }\textbf {\bibinfo {volume}
  {14}},\ \bibinfo {pages} {5371} (\bibinfo {year} {2014})}\BibitemShut
  {NoStop}%
\bibitem [{\citenamefont {Massicotte}\ \emph {et~al.}(2021)\citenamefont
  {Massicotte}, \citenamefont {Soavi}, \citenamefont {Principi},\ and\
  \citenamefont {Tielrooij}}]{massicotte2021hot}%
  \BibitemOpen
  \bibfield  {author} {\bibinfo {author} {\bibfnamefont {M.}~\bibnamefont
  {Massicotte}}, \bibinfo {author} {\bibfnamefont {G.}~\bibnamefont {Soavi}},
  \bibinfo {author} {\bibfnamefont {A.}~\bibnamefont {Principi}}, \ and\
  \bibinfo {author} {\bibfnamefont {K.-J.}\ \bibnamefont {Tielrooij}},\
  }\href@noop {} {\bibfield  {journal} {\bibinfo  {journal} {Nanoscale}\
  }\textbf {\bibinfo {volume} {13}},\ \bibinfo {pages} {8376} (\bibinfo {year}
  {2021})}\BibitemShut {NoStop}%
\bibitem [{\citenamefont {Gierz}\ \emph {et~al.}(2013)\citenamefont {Gierz},
  \citenamefont {Petersen}, \citenamefont {Mitrano}, \citenamefont {Cacho},
  \citenamefont {Turcu}, \citenamefont {Springate}, \citenamefont {St{\"o}hr},
  \citenamefont {K{\"o}hler}, \citenamefont {Starke},\ and\ \citenamefont
  {Cavalleri}}]{gierz2013snapshots}%
  \BibitemOpen
  \bibfield  {author} {\bibinfo {author} {\bibfnamefont {I.}~\bibnamefont
  {Gierz}}, \bibinfo {author} {\bibfnamefont {J.~C.}\ \bibnamefont {Petersen}},
  \bibinfo {author} {\bibfnamefont {M.}~\bibnamefont {Mitrano}}, \bibinfo
  {author} {\bibfnamefont {C.}~\bibnamefont {Cacho}}, \bibinfo {author}
  {\bibfnamefont {I.~E.}\ \bibnamefont {Turcu}}, \bibinfo {author}
  {\bibfnamefont {E.}~\bibnamefont {Springate}}, \bibinfo {author}
  {\bibfnamefont {A.}~\bibnamefont {St{\"o}hr}}, \bibinfo {author}
  {\bibfnamefont {A.}~\bibnamefont {K{\"o}hler}}, \bibinfo {author}
  {\bibfnamefont {U.}~\bibnamefont {Starke}}, \ and\ \bibinfo {author}
  {\bibfnamefont {A.}~\bibnamefont {Cavalleri}},\ }\href@noop {} {\bibfield
  {journal} {\bibinfo  {journal} {Nature materials}\ }\textbf {\bibinfo
  {volume} {12}},\ \bibinfo {pages} {1119} (\bibinfo {year}
  {2013})}\BibitemShut {NoStop}%
\bibitem [{Note1()}]{Note1}%
  \BibitemOpen
  \bibinfo {note} {See section \ref {results} and Appendix \ref
  {Section:Transport_and_PC_measurements}\ref {Subsection:TIA} for detailed
  discussions. Also see Ref.~\cite {gazzano2019observation}}\BibitemShut
  {NoStop}%
\bibitem [{\citenamefont {Horowitz}\ and\ \citenamefont
  {Hill}(2002)}]{horowitz2002art}%
  \BibitemOpen
  \bibfield  {author} {\bibinfo {author} {\bibfnamefont {P.}~\bibnamefont
  {Horowitz}}\ and\ \bibinfo {author} {\bibfnamefont {W.}~\bibnamefont
  {Hill}},\ }\href@noop {} {\emph {\bibinfo {title} {The art of electronics}}}\
  (\bibinfo  {publisher} {Cambridge university press Cambridge},\ \bibinfo
  {year} {2002})\BibitemShut {NoStop}%
\bibitem [{\citenamefont {McIver}\ \emph {et~al.}(2020)\citenamefont {McIver},
  \citenamefont {Schulte}, \citenamefont {Stein}, \citenamefont {Matsuyama},
  \citenamefont {Jotzu}, \citenamefont {Meier},\ and\ \citenamefont
  {Cavalleri}}]{mciver2020light}%
  \BibitemOpen
  \bibfield  {author} {\bibinfo {author} {\bibfnamefont {J.~W.}\ \bibnamefont
  {McIver}}, \bibinfo {author} {\bibfnamefont {B.}~\bibnamefont {Schulte}},
  \bibinfo {author} {\bibfnamefont {F.-U.}\ \bibnamefont {Stein}}, \bibinfo
  {author} {\bibfnamefont {T.}~\bibnamefont {Matsuyama}}, \bibinfo {author}
  {\bibfnamefont {G.}~\bibnamefont {Jotzu}}, \bibinfo {author} {\bibfnamefont
  {G.}~\bibnamefont {Meier}}, \ and\ \bibinfo {author} {\bibfnamefont
  {A.}~\bibnamefont {Cavalleri}},\ }\href@noop {} {\bibfield  {journal}
  {\bibinfo  {journal} {Nature physics}\ }\textbf {\bibinfo {volume} {16}},\
  \bibinfo {pages} {38} (\bibinfo {year} {2020})}\BibitemShut {NoStop}%
\bibitem [{\citenamefont {Donati}(2021)}]{donati2021photodetectors}%
  \BibitemOpen
  \bibfield  {author} {\bibinfo {author} {\bibfnamefont {S.}~\bibnamefont
  {Donati}},\ }\href@noop {} {\emph {\bibinfo {title} {Photodetectors: devices,
  circuits and applications}}}\ (\bibinfo  {publisher} {John Wiley \& Sons},\
  \bibinfo {year} {2021})\BibitemShut {NoStop}%
\bibitem [{\citenamefont {Song}\ and\ \citenamefont
  {Levitov}(2014)}]{song2014shockley}%
  \BibitemOpen
  \bibfield  {author} {\bibinfo {author} {\bibfnamefont {J.~C.}\ \bibnamefont
  {Song}}\ and\ \bibinfo {author} {\bibfnamefont {L.~S.}\ \bibnamefont
  {Levitov}},\ }\href@noop {} {\bibfield  {journal} {\bibinfo  {journal}
  {Physical Review B}\ }\textbf {\bibinfo {volume} {90}},\ \bibinfo {pages}
  {075415} (\bibinfo {year} {2014})}\BibitemShut {NoStop}%
\bibitem [{\citenamefont {Winzer}\ \emph {et~al.}(2017)\citenamefont {Winzer},
  \citenamefont {Mittendorff}, \citenamefont {Winnerl}, \citenamefont
  {Mittenzwey}, \citenamefont {Jago}, \citenamefont {Helm}, \citenamefont
  {Malic},\ and\ \citenamefont {Knorr}}]{winzer2017unconventional}%
  \BibitemOpen
  \bibfield  {author} {\bibinfo {author} {\bibfnamefont {T.}~\bibnamefont
  {Winzer}}, \bibinfo {author} {\bibfnamefont {M.}~\bibnamefont {Mittendorff}},
  \bibinfo {author} {\bibfnamefont {S.}~\bibnamefont {Winnerl}}, \bibinfo
  {author} {\bibfnamefont {H.}~\bibnamefont {Mittenzwey}}, \bibinfo {author}
  {\bibfnamefont {R.}~\bibnamefont {Jago}}, \bibinfo {author} {\bibfnamefont
  {M.}~\bibnamefont {Helm}}, \bibinfo {author} {\bibfnamefont {E.}~\bibnamefont
  {Malic}}, \ and\ \bibinfo {author} {\bibfnamefont {A.}~\bibnamefont
  {Knorr}},\ }\href@noop {} {\bibfield  {journal} {\bibinfo  {journal} {Nature
  communications}\ }\textbf {\bibinfo {volume} {8}},\ \bibinfo {pages} {1}
  (\bibinfo {year} {2017})}\BibitemShut {NoStop}%
\bibitem [{\citenamefont {Shen}(1984)}]{shen1984principles}%
  \BibitemOpen
  \bibfield  {author} {\bibinfo {author} {\bibfnamefont {Y.-R.}\ \bibnamefont
  {Shen}},\ }\href@noop {} {\emph {\bibinfo {title} {The principles of
  nonlinear optics}}}\ (\bibinfo {year} {1984})\BibitemShut {NoStop}%
\bibitem [{\citenamefont {Sonntag}\ \emph {et~al.}(2018)\citenamefont
  {Sonntag}, \citenamefont {Reichardt}, \citenamefont {Wirtz}, \citenamefont
  {Beschoten}, \citenamefont {Katsnelson}, \citenamefont {Libisch},\ and\
  \citenamefont {Stampfer}}]{sonntag2018impact}%
  \BibitemOpen
  \bibfield  {author} {\bibinfo {author} {\bibfnamefont {J.}~\bibnamefont
  {Sonntag}}, \bibinfo {author} {\bibfnamefont {S.}~\bibnamefont {Reichardt}},
  \bibinfo {author} {\bibfnamefont {L.}~\bibnamefont {Wirtz}}, \bibinfo
  {author} {\bibfnamefont {B.}~\bibnamefont {Beschoten}}, \bibinfo {author}
  {\bibfnamefont {M.~I.}\ \bibnamefont {Katsnelson}}, \bibinfo {author}
  {\bibfnamefont {F.}~\bibnamefont {Libisch}}, \ and\ \bibinfo {author}
  {\bibfnamefont {C.}~\bibnamefont {Stampfer}},\ }\href@noop {} {\bibfield
  {journal} {\bibinfo  {journal} {Physical review letters}\ }\textbf {\bibinfo
  {volume} {120}},\ \bibinfo {pages} {187701} (\bibinfo {year}
  {2018})}\BibitemShut {NoStop}%
\bibitem [{\citenamefont {Shockley}(1938)}]{shockley1938currents}%
  \BibitemOpen
  \bibfield  {author} {\bibinfo {author} {\bibfnamefont {W.}~\bibnamefont
  {Shockley}},\ }\href@noop {} {\bibfield  {journal} {\bibinfo  {journal}
  {Journal of applied physics}\ }\textbf {\bibinfo {volume} {9}},\ \bibinfo
  {pages} {635} (\bibinfo {year} {1938})}\BibitemShut {NoStop}%
\bibitem [{\citenamefont {Ramo}(1939)}]{ramo1939currents}%
  \BibitemOpen
  \bibfield  {author} {\bibinfo {author} {\bibfnamefont {S.}~\bibnamefont
  {Ramo}},\ }\href@noop {} {\bibfield  {journal} {\bibinfo  {journal}
  {Proceedings of the IRE}\ }\textbf {\bibinfo {volume} {27}},\ \bibinfo
  {pages} {584} (\bibinfo {year} {1939})}\BibitemShut {NoStop}%
\bibitem [{Note2()}]{Note2}%
  \BibitemOpen
  \bibinfo {note} {The number of resolvable LLs is $\sim $ 29 and this number
  does not change with B~\cite {orlita2011carrier}. We limit to 21 due to
  computational cost}\BibitemShut {NoStop}%
\bibitem [{\citenamefont {Lee}\ \emph {et~al.}(2017)\citenamefont {Lee},
  \citenamefont {Huang}, \citenamefont {Efetov}, \citenamefont {Wei},
  \citenamefont {Hart}, \citenamefont {Taniguchi}, \citenamefont {Watanabe},
  \citenamefont {Yacoby},\ and\ \citenamefont {Kim}}]{lee2017inducing}%
  \BibitemOpen
  \bibfield  {author} {\bibinfo {author} {\bibfnamefont {G.-H.}\ \bibnamefont
  {Lee}}, \bibinfo {author} {\bibfnamefont {K.-F.}\ \bibnamefont {Huang}},
  \bibinfo {author} {\bibfnamefont {D.~K.}\ \bibnamefont {Efetov}}, \bibinfo
  {author} {\bibfnamefont {D.~S.}\ \bibnamefont {Wei}}, \bibinfo {author}
  {\bibfnamefont {S.}~\bibnamefont {Hart}}, \bibinfo {author} {\bibfnamefont
  {T.}~\bibnamefont {Taniguchi}}, \bibinfo {author} {\bibfnamefont
  {K.}~\bibnamefont {Watanabe}}, \bibinfo {author} {\bibfnamefont
  {A.}~\bibnamefont {Yacoby}}, \ and\ \bibinfo {author} {\bibfnamefont
  {P.}~\bibnamefont {Kim}},\ }\href@noop {} {\bibfield  {journal} {\bibinfo
  {journal} {Nature Physics}\ }\textbf {\bibinfo {volume} {13}},\ \bibinfo
  {pages} {693} (\bibinfo {year} {2017})}\BibitemShut {NoStop}%
\bibitem [{\citenamefont {Shalom}\ \emph {et~al.}(2016)\citenamefont {Shalom},
  \citenamefont {Zhu}, \citenamefont {Fal’Ko}, \citenamefont {Mishchenko},
  \citenamefont {Kretinin}, \citenamefont {Novoselov}, \citenamefont {Woods},
  \citenamefont {Watanabe}, \citenamefont {Taniguchi}, \citenamefont {Geim}
  \emph {et~al.}}]{shalom2016quantum}%
  \BibitemOpen
  \bibfield  {author} {\bibinfo {author} {\bibfnamefont {M.~B.}\ \bibnamefont
  {Shalom}}, \bibinfo {author} {\bibfnamefont {M.}~\bibnamefont {Zhu}},
  \bibinfo {author} {\bibfnamefont {V.}~\bibnamefont {Fal’Ko}}, \bibinfo
  {author} {\bibfnamefont {A.}~\bibnamefont {Mishchenko}}, \bibinfo {author}
  {\bibfnamefont {A.}~\bibnamefont {Kretinin}}, \bibinfo {author}
  {\bibfnamefont {K.}~\bibnamefont {Novoselov}}, \bibinfo {author}
  {\bibfnamefont {C.}~\bibnamefont {Woods}}, \bibinfo {author} {\bibfnamefont
  {K.}~\bibnamefont {Watanabe}}, \bibinfo {author} {\bibfnamefont
  {T.}~\bibnamefont {Taniguchi}}, \bibinfo {author} {\bibfnamefont
  {A.}~\bibnamefont {Geim}},  \emph {et~al.},\ }\href@noop {} {\bibfield
  {journal} {\bibinfo  {journal} {Nature Physics}\ }\textbf {\bibinfo {volume}
  {12}},\ \bibinfo {pages} {318} (\bibinfo {year} {2016})}\BibitemShut
  {NoStop}%
\bibitem [{\citenamefont {Hunt}\ \emph {et~al.}(2013)\citenamefont {Hunt},
  \citenamefont {Sanchez-Yamagishi}, \citenamefont {Young}, \citenamefont
  {Yankowitz}, \citenamefont {LeRoy}, \citenamefont {Watanabe}, \citenamefont
  {Taniguchi}, \citenamefont {Moon}, \citenamefont {Koshino}, \citenamefont
  {Jarillo-Herrero} \emph {et~al.}}]{hunt2013massive}%
  \BibitemOpen
  \bibfield  {author} {\bibinfo {author} {\bibfnamefont {B.}~\bibnamefont
  {Hunt}}, \bibinfo {author} {\bibfnamefont {J.~D.}\ \bibnamefont
  {Sanchez-Yamagishi}}, \bibinfo {author} {\bibfnamefont {A.~F.}\ \bibnamefont
  {Young}}, \bibinfo {author} {\bibfnamefont {M.}~\bibnamefont {Yankowitz}},
  \bibinfo {author} {\bibfnamefont {B.~J.}\ \bibnamefont {LeRoy}}, \bibinfo
  {author} {\bibfnamefont {K.}~\bibnamefont {Watanabe}}, \bibinfo {author}
  {\bibfnamefont {T.}~\bibnamefont {Taniguchi}}, \bibinfo {author}
  {\bibfnamefont {P.}~\bibnamefont {Moon}}, \bibinfo {author} {\bibfnamefont
  {M.}~\bibnamefont {Koshino}}, \bibinfo {author} {\bibfnamefont
  {P.}~\bibnamefont {Jarillo-Herrero}},  \emph {et~al.},\ }\href@noop {}
  {\bibfield  {journal} {\bibinfo  {journal} {Science}\ }\textbf {\bibinfo
  {volume} {340}},\ \bibinfo {pages} {1427} (\bibinfo {year}
  {2013})}\BibitemShut {NoStop}%
\bibitem [{\citenamefont {Yang}\ \emph {et~al.}(2021)\citenamefont {Yang},
  \citenamefont {Zibrov}, \citenamefont {Bai}, \citenamefont {Taniguchi},
  \citenamefont {Watanabe}, \citenamefont {Zaletel},\ and\ \citenamefont
  {Young}}]{yang2021experimental}%
  \BibitemOpen
  \bibfield  {author} {\bibinfo {author} {\bibfnamefont {F.}~\bibnamefont
  {Yang}}, \bibinfo {author} {\bibfnamefont {A.~A.}\ \bibnamefont {Zibrov}},
  \bibinfo {author} {\bibfnamefont {R.}~\bibnamefont {Bai}}, \bibinfo {author}
  {\bibfnamefont {T.}~\bibnamefont {Taniguchi}}, \bibinfo {author}
  {\bibfnamefont {K.}~\bibnamefont {Watanabe}}, \bibinfo {author}
  {\bibfnamefont {M.~P.}\ \bibnamefont {Zaletel}}, \ and\ \bibinfo {author}
  {\bibfnamefont {A.~F.}\ \bibnamefont {Young}},\ }\href@noop {} {\bibfield
  {journal} {\bibinfo  {journal} {Physical Review Letters}\ }\textbf {\bibinfo
  {volume} {126}},\ \bibinfo {pages} {156802} (\bibinfo {year}
  {2021})}\BibitemShut {NoStop}%
\bibitem [{\citenamefont {Zhang}\ \emph {et~al.}(2006)\citenamefont {Zhang},
  \citenamefont {Jiang}, \citenamefont {Small}, \citenamefont {Purewal},
  \citenamefont {Tan}, \citenamefont {Fazlollahi}, \citenamefont {Chudow},
  \citenamefont {Jaszczak}, \citenamefont {Stormer},\ and\ \citenamefont
  {Kim}}]{zhang2006landau}%
  \BibitemOpen
  \bibfield  {author} {\bibinfo {author} {\bibfnamefont {Y.}~\bibnamefont
  {Zhang}}, \bibinfo {author} {\bibfnamefont {Z.}~\bibnamefont {Jiang}},
  \bibinfo {author} {\bibfnamefont {J.}~\bibnamefont {Small}}, \bibinfo
  {author} {\bibfnamefont {M.}~\bibnamefont {Purewal}}, \bibinfo {author}
  {\bibfnamefont {Y.-W.}\ \bibnamefont {Tan}}, \bibinfo {author} {\bibfnamefont
  {M.}~\bibnamefont {Fazlollahi}}, \bibinfo {author} {\bibfnamefont
  {J.}~\bibnamefont {Chudow}}, \bibinfo {author} {\bibfnamefont
  {J.}~\bibnamefont {Jaszczak}}, \bibinfo {author} {\bibfnamefont
  {H.}~\bibnamefont {Stormer}}, \ and\ \bibinfo {author} {\bibfnamefont
  {P.}~\bibnamefont {Kim}},\ }\href@noop {} {\bibfield  {journal} {\bibinfo
  {journal} {Physical review letters}\ }\textbf {\bibinfo {volume} {96}},\
  \bibinfo {pages} {136806} (\bibinfo {year} {2006})}\BibitemShut {NoStop}%
\bibitem [{\citenamefont {Young}\ \emph {et~al.}(2012)\citenamefont {Young},
  \citenamefont {Dean}, \citenamefont {Wang}, \citenamefont {Ren},
  \citenamefont {Cadden-Zimansky}, \citenamefont {Watanabe}, \citenamefont
  {Taniguchi}, \citenamefont {Hone}, \citenamefont {Shepard},\ and\
  \citenamefont {Kim}}]{young2012spin}%
  \BibitemOpen
  \bibfield  {author} {\bibinfo {author} {\bibfnamefont {A.~F.}\ \bibnamefont
  {Young}}, \bibinfo {author} {\bibfnamefont {C.~R.}\ \bibnamefont {Dean}},
  \bibinfo {author} {\bibfnamefont {L.}~\bibnamefont {Wang}}, \bibinfo {author}
  {\bibfnamefont {H.}~\bibnamefont {Ren}}, \bibinfo {author} {\bibfnamefont
  {P.}~\bibnamefont {Cadden-Zimansky}}, \bibinfo {author} {\bibfnamefont
  {K.}~\bibnamefont {Watanabe}}, \bibinfo {author} {\bibfnamefont
  {T.}~\bibnamefont {Taniguchi}}, \bibinfo {author} {\bibfnamefont
  {J.}~\bibnamefont {Hone}}, \bibinfo {author} {\bibfnamefont {K.~L.}\
  \bibnamefont {Shepard}}, \ and\ \bibinfo {author} {\bibfnamefont
  {P.}~\bibnamefont {Kim}},\ }\href@noop {} {\bibfield  {journal} {\bibinfo
  {journal} {Nature Physics}\ }\textbf {\bibinfo {volume} {8}},\ \bibinfo
  {pages} {550} (\bibinfo {year} {2012})}\BibitemShut {NoStop}%
\bibitem [{\citenamefont {Amet}\ \emph {et~al.}(2013)\citenamefont {Amet},
  \citenamefont {Williams}, \citenamefont {Watanabe}, \citenamefont
  {Taniguchi},\ and\ \citenamefont {Goldhaber-Gordon}}]{amet2013insulating}%
  \BibitemOpen
  \bibfield  {author} {\bibinfo {author} {\bibfnamefont {F.}~\bibnamefont
  {Amet}}, \bibinfo {author} {\bibfnamefont {J.}~\bibnamefont {Williams}},
  \bibinfo {author} {\bibfnamefont {K.}~\bibnamefont {Watanabe}}, \bibinfo
  {author} {\bibfnamefont {T.}~\bibnamefont {Taniguchi}}, \ and\ \bibinfo
  {author} {\bibfnamefont {D.}~\bibnamefont {Goldhaber-Gordon}},\ }\href@noop
  {} {\bibfield  {journal} {\bibinfo  {journal} {Physical review letters}\
  }\textbf {\bibinfo {volume} {110}},\ \bibinfo {pages} {216601} (\bibinfo
  {year} {2013})}\BibitemShut {NoStop}%
\bibitem [{\citenamefont {Pogna}\ \emph {et~al.}(0)\citenamefont {Pogna},
  \citenamefont {Jia}, \citenamefont {Principi}, \citenamefont {Block},
  \citenamefont {Banszerus}, \citenamefont {Zhang}, \citenamefont {Liu},
  \citenamefont {Sohier}, \citenamefont {Forti}, \citenamefont
  {Soundarapandian}, \citenamefont {Terrés}, \citenamefont {Mehew},
  \citenamefont {Trovatello}, \citenamefont {Coletti}, \citenamefont {Koppens},
  \citenamefont {Bonn}, \citenamefont {Wang}, \citenamefont {van Hulst},
  \citenamefont {Verstraete}, \citenamefont {Peng}, \citenamefont {Liu},
  \citenamefont {Stampfer}, \citenamefont {Cerullo},\ and\ \citenamefont
  {Tielrooij}}]{pogna2021hot}%
  \BibitemOpen
  \bibfield  {author} {\bibinfo {author} {\bibfnamefont {E.~A.~A.}\
  \bibnamefont {Pogna}}, \bibinfo {author} {\bibfnamefont {X.}~\bibnamefont
  {Jia}}, \bibinfo {author} {\bibfnamefont {A.}~\bibnamefont {Principi}},
  \bibinfo {author} {\bibfnamefont {A.}~\bibnamefont {Block}}, \bibinfo
  {author} {\bibfnamefont {L.}~\bibnamefont {Banszerus}}, \bibinfo {author}
  {\bibfnamefont {J.}~\bibnamefont {Zhang}}, \bibinfo {author} {\bibfnamefont
  {X.}~\bibnamefont {Liu}}, \bibinfo {author} {\bibfnamefont {T.}~\bibnamefont
  {Sohier}}, \bibinfo {author} {\bibfnamefont {S.}~\bibnamefont {Forti}},
  \bibinfo {author} {\bibfnamefont {K.}~\bibnamefont {Soundarapandian}},
  \bibinfo {author} {\bibfnamefont {B.}~\bibnamefont {Terrés}}, \bibinfo
  {author} {\bibfnamefont {J.~D.}\ \bibnamefont {Mehew}}, \bibinfo {author}
  {\bibfnamefont {C.}~\bibnamefont {Trovatello}}, \bibinfo {author}
  {\bibfnamefont {C.}~\bibnamefont {Coletti}}, \bibinfo {author} {\bibfnamefont
  {F.~H.~L.}\ \bibnamefont {Koppens}}, \bibinfo {author} {\bibfnamefont
  {M.}~\bibnamefont {Bonn}}, \bibinfo {author} {\bibfnamefont {H.~I.}\
  \bibnamefont {Wang}}, \bibinfo {author} {\bibfnamefont {N.}~\bibnamefont {van
  Hulst}}, \bibinfo {author} {\bibfnamefont {M.~J.}\ \bibnamefont
  {Verstraete}}, \bibinfo {author} {\bibfnamefont {H.}~\bibnamefont {Peng}},
  \bibinfo {author} {\bibfnamefont {Z.}~\bibnamefont {Liu}}, \bibinfo {author}
  {\bibfnamefont {C.}~\bibnamefont {Stampfer}}, \bibinfo {author}
  {\bibfnamefont {G.}~\bibnamefont {Cerullo}}, \ and\ \bibinfo {author}
  {\bibfnamefont {K.-J.}\ \bibnamefont {Tielrooij}},\ }\href {\doibase
  10.1021/acsnano.0c10864} {\bibfield  {journal} {\bibinfo  {journal} {ACS
  Nano}\ }\textbf {\bibinfo {volume} {0}},\ \bibinfo {pages} {null} (\bibinfo
  {year} {0})},\ \bibinfo {note} {pMID: 34139125},\ \Eprint
  {http://arxiv.org/abs/https://doi.org/10.1021/acsnano.0c10864}
  {https://doi.org/10.1021/acsnano.0c10864} \BibitemShut {NoStop}%
\bibitem [{\citenamefont {Walsh}\ \emph {et~al.}(2021)\citenamefont {Walsh},
  \citenamefont {Jung}, \citenamefont {Lee}, \citenamefont {Efetov},
  \citenamefont {Wu}, \citenamefont {Huang}, \citenamefont {Ohki},
  \citenamefont {Taniguchi}, \citenamefont {Watanabe}, \citenamefont {Kim}
  \emph {et~al.}}]{walsh2021josephson}%
  \BibitemOpen
  \bibfield  {author} {\bibinfo {author} {\bibfnamefont {E.~D.}\ \bibnamefont
  {Walsh}}, \bibinfo {author} {\bibfnamefont {W.}~\bibnamefont {Jung}},
  \bibinfo {author} {\bibfnamefont {G.-H.}\ \bibnamefont {Lee}}, \bibinfo
  {author} {\bibfnamefont {D.~K.}\ \bibnamefont {Efetov}}, \bibinfo {author}
  {\bibfnamefont {B.-I.}\ \bibnamefont {Wu}}, \bibinfo {author} {\bibfnamefont
  {K.-F.}\ \bibnamefont {Huang}}, \bibinfo {author} {\bibfnamefont {T.~A.}\
  \bibnamefont {Ohki}}, \bibinfo {author} {\bibfnamefont {T.}~\bibnamefont
  {Taniguchi}}, \bibinfo {author} {\bibfnamefont {K.}~\bibnamefont {Watanabe}},
  \bibinfo {author} {\bibfnamefont {P.}~\bibnamefont {Kim}},  \emph {et~al.},\
  }\href@noop {} {\bibfield  {journal} {\bibinfo  {journal} {Science}\ }\textbf
  {\bibinfo {volume} {372}},\ \bibinfo {pages} {409} (\bibinfo {year}
  {2021})}\BibitemShut {NoStop}%
\bibitem [{\citenamefont {Tielrooij}\ \emph {et~al.}(2015)\citenamefont
  {Tielrooij}, \citenamefont {Piatkowski}, \citenamefont {Massicotte},
  \citenamefont {Woessner}, \citenamefont {Ma}, \citenamefont {Lee},
  \citenamefont {Myhro}, \citenamefont {Lau}, \citenamefont {Jarillo-Herrero},
  \citenamefont {van Hulst} \emph {et~al.}}]{tielrooij2015generation}%
  \BibitemOpen
  \bibfield  {author} {\bibinfo {author} {\bibfnamefont {K.-J.}\ \bibnamefont
  {Tielrooij}}, \bibinfo {author} {\bibfnamefont {L.}~\bibnamefont
  {Piatkowski}}, \bibinfo {author} {\bibfnamefont {M.}~\bibnamefont
  {Massicotte}}, \bibinfo {author} {\bibfnamefont {A.}~\bibnamefont
  {Woessner}}, \bibinfo {author} {\bibfnamefont {Q.}~\bibnamefont {Ma}},
  \bibinfo {author} {\bibfnamefont {Y.}~\bibnamefont {Lee}}, \bibinfo {author}
  {\bibfnamefont {K.~S.}\ \bibnamefont {Myhro}}, \bibinfo {author}
  {\bibfnamefont {C.~N.}\ \bibnamefont {Lau}}, \bibinfo {author} {\bibfnamefont
  {P.}~\bibnamefont {Jarillo-Herrero}}, \bibinfo {author} {\bibfnamefont
  {N.~F.}\ \bibnamefont {van Hulst}},  \emph {et~al.},\ }\href@noop {}
  {\bibfield  {journal} {\bibinfo  {journal} {Nature nanotechnology}\ }\textbf
  {\bibinfo {volume} {10}},\ \bibinfo {pages} {437} (\bibinfo {year}
  {2015})}\BibitemShut {NoStop}%
\bibitem [{\citenamefont {Koppens}\ \emph {et~al.}(2014)\citenamefont
  {Koppens}, \citenamefont {Mueller}, \citenamefont {Avouris}, \citenamefont
  {Ferrari}, \citenamefont {Vitiello},\ and\ \citenamefont
  {Polini}}]{koppens2014photodetectors}%
  \BibitemOpen
  \bibfield  {author} {\bibinfo {author} {\bibfnamefont {F.}~\bibnamefont
  {Koppens}}, \bibinfo {author} {\bibfnamefont {T.}~\bibnamefont {Mueller}},
  \bibinfo {author} {\bibfnamefont {P.}~\bibnamefont {Avouris}}, \bibinfo
  {author} {\bibfnamefont {A.}~\bibnamefont {Ferrari}}, \bibinfo {author}
  {\bibfnamefont {M.}~\bibnamefont {Vitiello}}, \ and\ \bibinfo {author}
  {\bibfnamefont {M.}~\bibnamefont {Polini}},\ }\href@noop {} {\bibfield
  {journal} {\bibinfo  {journal} {Nature nanotechnology}\ }\textbf {\bibinfo
  {volume} {9}},\ \bibinfo {pages} {780} (\bibinfo {year} {2014})}\BibitemShut
  {NoStop}%
\bibitem [{\citenamefont {Gabor}\ \emph {et~al.}(2011)\citenamefont {Gabor},
  \citenamefont {Song}, \citenamefont {Ma}, \citenamefont {Nair}, \citenamefont
  {Taychatanapat}, \citenamefont {Watanabe}, \citenamefont {Taniguchi},
  \citenamefont {Levitov},\ and\ \citenamefont
  {Jarillo-Herrero}}]{gabor2011hot}%
  \BibitemOpen
  \bibfield  {author} {\bibinfo {author} {\bibfnamefont {N.~M.}\ \bibnamefont
  {Gabor}}, \bibinfo {author} {\bibfnamefont {J.~C.}\ \bibnamefont {Song}},
  \bibinfo {author} {\bibfnamefont {Q.}~\bibnamefont {Ma}}, \bibinfo {author}
  {\bibfnamefont {N.~L.}\ \bibnamefont {Nair}}, \bibinfo {author}
  {\bibfnamefont {T.}~\bibnamefont {Taychatanapat}}, \bibinfo {author}
  {\bibfnamefont {K.}~\bibnamefont {Watanabe}}, \bibinfo {author}
  {\bibfnamefont {T.}~\bibnamefont {Taniguchi}}, \bibinfo {author}
  {\bibfnamefont {L.~S.}\ \bibnamefont {Levitov}}, \ and\ \bibinfo {author}
  {\bibfnamefont {P.}~\bibnamefont {Jarillo-Herrero}},\ }\href@noop {}
  {\bibfield  {journal} {\bibinfo  {journal} {Science}\ }\textbf {\bibinfo
  {volume} {334}},\ \bibinfo {pages} {648} (\bibinfo {year}
  {2011})}\BibitemShut {NoStop}%
\bibitem [{\citenamefont {Ma}\ \emph {et~al.}(2019)\citenamefont {Ma},
  \citenamefont {Lui}, \citenamefont {Song}, \citenamefont {Lin}, \citenamefont
  {Kong}, \citenamefont {Cao}, \citenamefont {Dinh}, \citenamefont {Nair},
  \citenamefont {Fang}, \citenamefont {Watanabe} \emph {et~al.}}]{ma2019giant}%
  \BibitemOpen
  \bibfield  {author} {\bibinfo {author} {\bibfnamefont {Q.}~\bibnamefont
  {Ma}}, \bibinfo {author} {\bibfnamefont {C.~H.}\ \bibnamefont {Lui}},
  \bibinfo {author} {\bibfnamefont {J.~C.}\ \bibnamefont {Song}}, \bibinfo
  {author} {\bibfnamefont {Y.}~\bibnamefont {Lin}}, \bibinfo {author}
  {\bibfnamefont {J.~F.}\ \bibnamefont {Kong}}, \bibinfo {author}
  {\bibfnamefont {Y.}~\bibnamefont {Cao}}, \bibinfo {author} {\bibfnamefont
  {T.~H.}\ \bibnamefont {Dinh}}, \bibinfo {author} {\bibfnamefont {N.~L.}\
  \bibnamefont {Nair}}, \bibinfo {author} {\bibfnamefont {W.}~\bibnamefont
  {Fang}}, \bibinfo {author} {\bibfnamefont {K.}~\bibnamefont {Watanabe}},
  \emph {et~al.},\ }\href@noop {} {\bibfield  {journal} {\bibinfo  {journal}
  {Nature nanotechnology}\ }\textbf {\bibinfo {volume} {14}},\ \bibinfo {pages}
  {145} (\bibinfo {year} {2019})}\BibitemShut {NoStop}%
\bibitem [{\citenamefont {Kang}\ \emph {et~al.}(2021)\citenamefont {Kang},
  \citenamefont {Sun}, \citenamefont {Luo}, \citenamefont {Lu}, \citenamefont
  {Chen}, \citenamefont {Kim}, \citenamefont {Jung}, \citenamefont {Gao},
  \citenamefont {Parluhutan}, \citenamefont {Ge} \emph
  {et~al.}}]{kang2021pseudo}%
  \BibitemOpen
  \bibfield  {author} {\bibinfo {author} {\bibfnamefont {D.-H.}\ \bibnamefont
  {Kang}}, \bibinfo {author} {\bibfnamefont {H.}~\bibnamefont {Sun}}, \bibinfo
  {author} {\bibfnamefont {M.}~\bibnamefont {Luo}}, \bibinfo {author}
  {\bibfnamefont {K.}~\bibnamefont {Lu}}, \bibinfo {author} {\bibfnamefont
  {M.}~\bibnamefont {Chen}}, \bibinfo {author} {\bibfnamefont {Y.}~\bibnamefont
  {Kim}}, \bibinfo {author} {\bibfnamefont {Y.}~\bibnamefont {Jung}}, \bibinfo
  {author} {\bibfnamefont {X.}~\bibnamefont {Gao}}, \bibinfo {author}
  {\bibfnamefont {S.~J.}\ \bibnamefont {Parluhutan}}, \bibinfo {author}
  {\bibfnamefont {J.}~\bibnamefont {Ge}},  \emph {et~al.},\ }\href@noop {}
  {\bibfield  {journal} {\bibinfo  {journal} {Nature Communications}\ }\textbf
  {\bibinfo {volume} {12}},\ \bibinfo {pages} {1} (\bibinfo {year}
  {2021})}\BibitemShut {NoStop}%
\bibitem [{\citenamefont {Han}\ \emph {et~al.}(2021)\citenamefont {Han},
  \citenamefont {Yang}, \citenamefont {Zhang}, \citenamefont {Wang},
  \citenamefont {Watanabe}, \citenamefont {Taniguchi}, \citenamefont {McEuen},\
  and\ \citenamefont {Ju}}]{han2021accurate}%
  \BibitemOpen
  \bibfield  {author} {\bibinfo {author} {\bibfnamefont {T.}~\bibnamefont
  {Han}}, \bibinfo {author} {\bibfnamefont {J.}~\bibnamefont {Yang}}, \bibinfo
  {author} {\bibfnamefont {Q.}~\bibnamefont {Zhang}}, \bibinfo {author}
  {\bibfnamefont {L.}~\bibnamefont {Wang}}, \bibinfo {author} {\bibfnamefont
  {K.}~\bibnamefont {Watanabe}}, \bibinfo {author} {\bibfnamefont
  {T.}~\bibnamefont {Taniguchi}}, \bibinfo {author} {\bibfnamefont {P.~L.}\
  \bibnamefont {McEuen}}, \ and\ \bibinfo {author} {\bibfnamefont
  {L.}~\bibnamefont {Ju}},\ }\href@noop {} {\bibfield  {journal} {\bibinfo
  {journal} {Physical Review Letters}\ }\textbf {\bibinfo {volume} {126}},\
  \bibinfo {pages} {146402} (\bibinfo {year} {2021})}\BibitemShut {NoStop}%
\bibitem [{\citenamefont {Cao}\ \emph {et~al.}(2021)\citenamefont {Cao},
  \citenamefont {Grass}, \citenamefont {Solomon},\ and\ \citenamefont
  {Hafezi}}]{PhysRevB.103.L241301}%
  \BibitemOpen
  \bibfield  {author} {\bibinfo {author} {\bibfnamefont {B.}~\bibnamefont
  {Cao}}, \bibinfo {author} {\bibfnamefont {T.}~\bibnamefont {Grass}}, \bibinfo
  {author} {\bibfnamefont {G.}~\bibnamefont {Solomon}}, \ and\ \bibinfo
  {author} {\bibfnamefont {M.}~\bibnamefont {Hafezi}},\ }\href {\doibase
  10.1103/PhysRevB.103.L241301} {\bibfield  {journal} {\bibinfo  {journal}
  {Phys. Rev. B}\ }\textbf {\bibinfo {volume} {103}},\ \bibinfo {pages}
  {L241301} (\bibinfo {year} {2021})}\BibitemShut {NoStop}%
\bibitem [{\citenamefont {But}\ \emph {et~al.}(2019)\citenamefont {But},
  \citenamefont {Mittendorff}, \citenamefont {Consejo}, \citenamefont {Teppe},
  \citenamefont {Mikhailov}, \citenamefont {Dvoretskii}, \citenamefont
  {Faugeras}, \citenamefont {Winnerl}, \citenamefont {Helm}, \citenamefont
  {Knap} \emph {et~al.}}]{but2019suppressed}%
  \BibitemOpen
  \bibfield  {author} {\bibinfo {author} {\bibfnamefont {D.}~\bibnamefont
  {But}}, \bibinfo {author} {\bibfnamefont {M.}~\bibnamefont {Mittendorff}},
  \bibinfo {author} {\bibfnamefont {C.}~\bibnamefont {Consejo}}, \bibinfo
  {author} {\bibfnamefont {F.}~\bibnamefont {Teppe}}, \bibinfo {author}
  {\bibfnamefont {N.}~\bibnamefont {Mikhailov}}, \bibinfo {author}
  {\bibfnamefont {S.}~\bibnamefont {Dvoretskii}}, \bibinfo {author}
  {\bibfnamefont {C.}~\bibnamefont {Faugeras}}, \bibinfo {author}
  {\bibfnamefont {S.}~\bibnamefont {Winnerl}}, \bibinfo {author} {\bibfnamefont
  {M.}~\bibnamefont {Helm}}, \bibinfo {author} {\bibfnamefont {W.}~\bibnamefont
  {Knap}},  \emph {et~al.},\ }\href@noop {} {\bibfield  {journal} {\bibinfo
  {journal} {Nature Photonics}\ }\textbf {\bibinfo {volume} {13}},\ \bibinfo
  {pages} {783} (\bibinfo {year} {2019})}\BibitemShut {NoStop}%
\bibitem [{\citenamefont {Abanin}\ \emph {et~al.}(2011)\citenamefont {Abanin},
  \citenamefont {Morozov}, \citenamefont {Ponomarenko}, \citenamefont
  {Gorbachev}, \citenamefont {Mayorov}, \citenamefont {Katsnelson},
  \citenamefont {Watanabe}, \citenamefont {Taniguchi}, \citenamefont
  {Novoselov}, \citenamefont {Levitov} \emph {et~al.}}]{abanin2011giant}%
  \BibitemOpen
  \bibfield  {author} {\bibinfo {author} {\bibfnamefont {D.}~\bibnamefont
  {Abanin}}, \bibinfo {author} {\bibfnamefont {S.}~\bibnamefont {Morozov}},
  \bibinfo {author} {\bibfnamefont {L.}~\bibnamefont {Ponomarenko}}, \bibinfo
  {author} {\bibfnamefont {R.}~\bibnamefont {Gorbachev}}, \bibinfo {author}
  {\bibfnamefont {A.}~\bibnamefont {Mayorov}}, \bibinfo {author} {\bibfnamefont
  {M.}~\bibnamefont {Katsnelson}}, \bibinfo {author} {\bibfnamefont
  {K.}~\bibnamefont {Watanabe}}, \bibinfo {author} {\bibfnamefont
  {T.}~\bibnamefont {Taniguchi}}, \bibinfo {author} {\bibfnamefont
  {K.}~\bibnamefont {Novoselov}}, \bibinfo {author} {\bibfnamefont
  {L.}~\bibnamefont {Levitov}},  \emph {et~al.},\ }\href@noop {} {\bibfield
  {journal} {\bibinfo  {journal} {Science}\ }\textbf {\bibinfo {volume}
  {332}},\ \bibinfo {pages} {328} (\bibinfo {year} {2011})}\BibitemShut
  {NoStop}%
\bibitem [{\citenamefont {Gorbachev}\ \emph {et~al.}(2014)\citenamefont
  {Gorbachev}, \citenamefont {Song}, \citenamefont {Yu}, \citenamefont
  {Kretinin}, \citenamefont {Withers}, \citenamefont {Cao}, \citenamefont
  {Mishchenko}, \citenamefont {Grigorieva}, \citenamefont {Novoselov},
  \citenamefont {Levitov} \emph {et~al.}}]{gorbachev2014detecting}%
  \BibitemOpen
  \bibfield  {author} {\bibinfo {author} {\bibfnamefont {R.}~\bibnamefont
  {Gorbachev}}, \bibinfo {author} {\bibfnamefont {J.}~\bibnamefont {Song}},
  \bibinfo {author} {\bibfnamefont {G.}~\bibnamefont {Yu}}, \bibinfo {author}
  {\bibfnamefont {A.}~\bibnamefont {Kretinin}}, \bibinfo {author}
  {\bibfnamefont {F.}~\bibnamefont {Withers}}, \bibinfo {author} {\bibfnamefont
  {Y.}~\bibnamefont {Cao}}, \bibinfo {author} {\bibfnamefont {A.}~\bibnamefont
  {Mishchenko}}, \bibinfo {author} {\bibfnamefont {I.}~\bibnamefont
  {Grigorieva}}, \bibinfo {author} {\bibfnamefont {K.~S.}\ \bibnamefont
  {Novoselov}}, \bibinfo {author} {\bibfnamefont {L.}~\bibnamefont {Levitov}},
  \emph {et~al.},\ }\href@noop {} {\bibfield  {journal} {\bibinfo  {journal}
  {Science}\ }\textbf {\bibinfo {volume} {346}},\ \bibinfo {pages} {448}
  (\bibinfo {year} {2014})}\BibitemShut {NoStop}%
\bibitem [{\citenamefont {Uri}\ \emph {et~al.}(2020)\citenamefont {Uri},
  \citenamefont {Kim}, \citenamefont {Bagani}, \citenamefont {Lewandowski},
  \citenamefont {Grover}, \citenamefont {Auerbach}, \citenamefont {Lachman},
  \citenamefont {Myasoedov}, \citenamefont {Taniguchi}, \citenamefont
  {Watanabe} \emph {et~al.}}]{uri2020nanoscale}%
  \BibitemOpen
  \bibfield  {author} {\bibinfo {author} {\bibfnamefont {A.}~\bibnamefont
  {Uri}}, \bibinfo {author} {\bibfnamefont {Y.}~\bibnamefont {Kim}}, \bibinfo
  {author} {\bibfnamefont {K.}~\bibnamefont {Bagani}}, \bibinfo {author}
  {\bibfnamefont {C.~K.}\ \bibnamefont {Lewandowski}}, \bibinfo {author}
  {\bibfnamefont {S.}~\bibnamefont {Grover}}, \bibinfo {author} {\bibfnamefont
  {N.}~\bibnamefont {Auerbach}}, \bibinfo {author} {\bibfnamefont {E.~O.}\
  \bibnamefont {Lachman}}, \bibinfo {author} {\bibfnamefont {Y.}~\bibnamefont
  {Myasoedov}}, \bibinfo {author} {\bibfnamefont {T.}~\bibnamefont
  {Taniguchi}}, \bibinfo {author} {\bibfnamefont {K.}~\bibnamefont {Watanabe}},
   \emph {et~al.},\ }\href@noop {} {\bibfield  {journal} {\bibinfo  {journal}
  {Nature Physics}\ }\textbf {\bibinfo {volume} {16}},\ \bibinfo {pages} {164}
  (\bibinfo {year} {2020})}\BibitemShut {NoStop}%
\bibitem [{\citenamefont {Aharon-Steinberg}\ \emph {et~al.}(2021)\citenamefont
  {Aharon-Steinberg}, \citenamefont {Marguerite}, \citenamefont {Perello},
  \citenamefont {Bagani}, \citenamefont {Holder}, \citenamefont {Myasoedov},
  \citenamefont {Levitov}, \citenamefont {Geim},\ and\ \citenamefont
  {Zeldov}}]{aharon2021long}%
  \BibitemOpen
  \bibfield  {author} {\bibinfo {author} {\bibfnamefont {A.}~\bibnamefont
  {Aharon-Steinberg}}, \bibinfo {author} {\bibfnamefont {A.}~\bibnamefont
  {Marguerite}}, \bibinfo {author} {\bibfnamefont {D.~J.}\ \bibnamefont
  {Perello}}, \bibinfo {author} {\bibfnamefont {K.}~\bibnamefont {Bagani}},
  \bibinfo {author} {\bibfnamefont {T.}~\bibnamefont {Holder}}, \bibinfo
  {author} {\bibfnamefont {Y.}~\bibnamefont {Myasoedov}}, \bibinfo {author}
  {\bibfnamefont {L.~S.}\ \bibnamefont {Levitov}}, \bibinfo {author}
  {\bibfnamefont {A.~K.}\ \bibnamefont {Geim}}, \ and\ \bibinfo {author}
  {\bibfnamefont {E.}~\bibnamefont {Zeldov}},\ }\href@noop {} {\bibfield
  {journal} {\bibinfo  {journal} {Nature}\ }\textbf {\bibinfo {volume} {593}},\
  \bibinfo {pages} {528} (\bibinfo {year} {2021})}\BibitemShut {NoStop}%
\bibitem [{\citenamefont {Malic}\ and\ \citenamefont
  {Knorr}(2013)}]{malic2013graphene}%
  \BibitemOpen
  \bibfield  {author} {\bibinfo {author} {\bibfnamefont {E.}~\bibnamefont
  {Malic}}\ and\ \bibinfo {author} {\bibfnamefont {A.}~\bibnamefont {Knorr}},\
  }\href@noop {} {\emph {\bibinfo {title} {Graphene and carbon nanotubes:
  ultrafast optics and relaxation dynamics}}}\ (\bibinfo  {publisher} {John
  Wiley \& Sons},\ \bibinfo {year} {2013})\BibitemShut {NoStop}%
\bibitem [{\citenamefont {Macdonald}(1994)}]{macdonald1994introduction}%
  \BibitemOpen
  \bibfield  {author} {\bibinfo {author} {\bibfnamefont {A.~H.}\ \bibnamefont
  {Macdonald}},\ }\href@noop {} {\bibfield  {journal} {\bibinfo  {journal}
  {arXiv preprint cond-mat/9410047}\ } (\bibinfo {year} {1994})}\BibitemShut
  {NoStop}%
\bibitem [{\citenamefont {Watanabe}\ \emph {et~al.}(2004)\citenamefont
  {Watanabe}, \citenamefont {Taniguchi},\ and\ \citenamefont
  {Kanda}}]{watanabe04}%
  \BibitemOpen
  \bibfield  {author} {\bibinfo {author} {\bibfnamefont {K.}~\bibnamefont
  {Watanabe}}, \bibinfo {author} {\bibfnamefont {T.}~\bibnamefont {Taniguchi}},
  \ and\ \bibinfo {author} {\bibfnamefont {H.}~\bibnamefont {Kanda}},\ }\href
  {\doibase 10.1038/nmat1134} {\bibfield  {journal} {\bibinfo  {journal} {Nat.
  Mater.}\ }\textbf {\bibinfo {volume} {3}},\ \bibinfo {pages} {404} (\bibinfo
  {year} {2004})}\BibitemShut {NoStop}%
\bibitem [{\citenamefont {Ferrari}\ \emph {et~al.}(2006)\citenamefont
  {Ferrari}, \citenamefont {Meyer}, \citenamefont {Scardaci}, \citenamefont
  {Casiraghi}, \citenamefont {Lazzeri}, \citenamefont {Mauri}, \citenamefont
  {Piscanec}, \citenamefont {Jiang}, \citenamefont {Novoselov}, \citenamefont
  {Roth},\ and\ \citenamefont {Geim}}]{ferrari06}%
  \BibitemOpen
  \bibfield  {author} {\bibinfo {author} {\bibfnamefont {A.~C.}\ \bibnamefont
  {Ferrari}}, \bibinfo {author} {\bibfnamefont {J.~C.}\ \bibnamefont {Meyer}},
  \bibinfo {author} {\bibfnamefont {V.}~\bibnamefont {Scardaci}}, \bibinfo
  {author} {\bibfnamefont {C.}~\bibnamefont {Casiraghi}}, \bibinfo {author}
  {\bibfnamefont {M.}~\bibnamefont {Lazzeri}}, \bibinfo {author} {\bibfnamefont
  {F.}~\bibnamefont {Mauri}}, \bibinfo {author} {\bibfnamefont
  {S.}~\bibnamefont {Piscanec}}, \bibinfo {author} {\bibfnamefont
  {D.}~\bibnamefont {Jiang}}, \bibinfo {author} {\bibfnamefont {K.~S.}\
  \bibnamefont {Novoselov}}, \bibinfo {author} {\bibfnamefont {S.}~\bibnamefont
  {Roth}}, \ and\ \bibinfo {author} {\bibfnamefont {A.~K.}\ \bibnamefont
  {Geim}},\ }\href {\doibase 10.1103/PhysRevLett.97.187401} {\bibfield
  {journal} {\bibinfo  {journal} {Phys. Rev. Lett.}\ }\textbf {\bibinfo
  {volume} {97}},\ \bibinfo {pages} {187401} (\bibinfo {year}
  {2006})}\BibitemShut {NoStop}%
\bibitem [{\citenamefont {Pizzocchero}\ \emph {et~al.}(2016)\citenamefont
  {Pizzocchero}, \citenamefont {Gammelgaard}, \citenamefont {Jessen},
  \citenamefont {Caridad}, \citenamefont {Wang}, \citenamefont {Hone},
  \citenamefont {B{\o}ggild},\ and\ \citenamefont {Booth}}]{pizzocchero16}%
  \BibitemOpen
  \bibfield  {author} {\bibinfo {author} {\bibfnamefont {F.}~\bibnamefont
  {Pizzocchero}}, \bibinfo {author} {\bibfnamefont {L.}~\bibnamefont
  {Gammelgaard}}, \bibinfo {author} {\bibfnamefont {B.~S.}\ \bibnamefont
  {Jessen}}, \bibinfo {author} {\bibfnamefont {J.~M.}\ \bibnamefont {Caridad}},
  \bibinfo {author} {\bibfnamefont {L.}~\bibnamefont {Wang}}, \bibinfo {author}
  {\bibfnamefont {J.}~\bibnamefont {Hone}}, \bibinfo {author} {\bibfnamefont
  {P.}~\bibnamefont {B{\o}ggild}}, \ and\ \bibinfo {author} {\bibfnamefont
  {T.~J.}\ \bibnamefont {Booth}},\ }\href {\doibase 10.1038/ncomms11894}
  {\bibfield  {journal} {\bibinfo  {journal} {Nat. Commun.}\ }\textbf {\bibinfo
  {volume} {7}},\ \bibinfo {pages} {11894} (\bibinfo {year}
  {2016})}\BibitemShut {NoStop}%
\bibitem [{\citenamefont {Wang}\ \emph {et~al.}(2013)\citenamefont {Wang},
  \citenamefont {Meric}, \citenamefont {Huang}, \citenamefont {Gao},
  \citenamefont {Gao}, \citenamefont {Tran}, \citenamefont {Taniguchi},
  \citenamefont {Watanabe}, \citenamefont {Campos}, \citenamefont {Muller},
  \citenamefont {Guo}, \citenamefont {Kim}, \citenamefont {Hone}, \citenamefont
  {Shepard},\ and\ \citenamefont {Dean}}]{wang13}%
  \BibitemOpen
  \bibfield  {author} {\bibinfo {author} {\bibfnamefont {L.}~\bibnamefont
  {Wang}}, \bibinfo {author} {\bibfnamefont {I.}~\bibnamefont {Meric}},
  \bibinfo {author} {\bibfnamefont {P.~Y.}\ \bibnamefont {Huang}}, \bibinfo
  {author} {\bibfnamefont {Q.}~\bibnamefont {Gao}}, \bibinfo {author}
  {\bibfnamefont {Y.}~\bibnamefont {Gao}}, \bibinfo {author} {\bibfnamefont
  {H.}~\bibnamefont {Tran}}, \bibinfo {author} {\bibfnamefont {T.}~\bibnamefont
  {Taniguchi}}, \bibinfo {author} {\bibfnamefont {K.}~\bibnamefont {Watanabe}},
  \bibinfo {author} {\bibfnamefont {L.~M.}\ \bibnamefont {Campos}}, \bibinfo
  {author} {\bibfnamefont {D.~A.}\ \bibnamefont {Muller}}, \bibinfo {author}
  {\bibfnamefont {J.}~\bibnamefont {Guo}}, \bibinfo {author} {\bibfnamefont
  {P.}~\bibnamefont {Kim}}, \bibinfo {author} {\bibfnamefont {J.}~\bibnamefont
  {Hone}}, \bibinfo {author} {\bibfnamefont {K.~L.}\ \bibnamefont {Shepard}}, \
  and\ \bibinfo {author} {\bibfnamefont {C.~R.}\ \bibnamefont {Dean}},\ }\href
  {\doibase 10.1126/science.1244358} {\bibfield  {journal} {\bibinfo  {journal}
  {Science}\ }\textbf {\bibinfo {volume} {342}},\ \bibinfo {pages} {614}
  (\bibinfo {year} {2013})}\BibitemShut {NoStop}%
\bibitem [{\citenamefont {Kim}\ \emph {et~al.}(2009)\citenamefont {Kim},
  \citenamefont {Nah}, \citenamefont {Jo}, \citenamefont {Shahrjerdi},
  \citenamefont {Colombo}, \citenamefont {Yao}, \citenamefont {Tutuc},\ and\
  \citenamefont {Banerjee}}]{kim2009realization}%
  \BibitemOpen
  \bibfield  {author} {\bibinfo {author} {\bibfnamefont {S.}~\bibnamefont
  {Kim}}, \bibinfo {author} {\bibfnamefont {J.}~\bibnamefont {Nah}}, \bibinfo
  {author} {\bibfnamefont {I.}~\bibnamefont {Jo}}, \bibinfo {author}
  {\bibfnamefont {D.}~\bibnamefont {Shahrjerdi}}, \bibinfo {author}
  {\bibfnamefont {L.}~\bibnamefont {Colombo}}, \bibinfo {author} {\bibfnamefont
  {Z.}~\bibnamefont {Yao}}, \bibinfo {author} {\bibfnamefont {E.}~\bibnamefont
  {Tutuc}}, \ and\ \bibinfo {author} {\bibfnamefont {S.~K.}\ \bibnamefont
  {Banerjee}},\ }\href@noop {} {\bibfield  {journal} {\bibinfo  {journal}
  {Applied Physics Letters}\ }\textbf {\bibinfo {volume} {94}},\ \bibinfo
  {pages} {062107} (\bibinfo {year} {2009})}\BibitemShut {NoStop}%
\bibitem [{\citenamefont {Abanin}\ and\ \citenamefont
  {Levitov}(2008)}]{abanin2008conformal}%
  \BibitemOpen
  \bibfield  {author} {\bibinfo {author} {\bibfnamefont {D.~A.}\ \bibnamefont
  {Abanin}}\ and\ \bibinfo {author} {\bibfnamefont {L.~S.}\ \bibnamefont
  {Levitov}},\ }\href@noop {} {\bibfield  {journal} {\bibinfo  {journal}
  {Physical Review B}\ }\textbf {\bibinfo {volume} {78}},\ \bibinfo {pages}
  {035416} (\bibinfo {year} {2008})}\BibitemShut {NoStop}%
\bibitem [{\citenamefont {Williams}\ \emph {et~al.}(2009)\citenamefont
  {Williams}, \citenamefont {Abanin}, \citenamefont {DiCarlo}, \citenamefont
  {Levitov},\ and\ \citenamefont {Marcus}}]{williams2009quantum}%
  \BibitemOpen
  \bibfield  {author} {\bibinfo {author} {\bibfnamefont {J.~R.}\ \bibnamefont
  {Williams}}, \bibinfo {author} {\bibfnamefont {D.~A.}\ \bibnamefont
  {Abanin}}, \bibinfo {author} {\bibfnamefont {L.}~\bibnamefont {DiCarlo}},
  \bibinfo {author} {\bibfnamefont {L.~S.}\ \bibnamefont {Levitov}}, \ and\
  \bibinfo {author} {\bibfnamefont {C.~M.}\ \bibnamefont {Marcus}},\
  }\href@noop {} {\bibfield  {journal} {\bibinfo  {journal} {Physical Review
  B}\ }\textbf {\bibinfo {volume} {80}},\ \bibinfo {pages} {045408} (\bibinfo
  {year} {2009})}\BibitemShut {NoStop}%
\bibitem [{\citenamefont {Sarma}\ \emph {et~al.}(2011)\citenamefont {Sarma},
  \citenamefont {Adam}, \citenamefont {Hwang},\ and\ \citenamefont
  {Rossi}}]{sarma2011electronic}%
  \BibitemOpen
  \bibfield  {author} {\bibinfo {author} {\bibfnamefont {S.~D.}\ \bibnamefont
  {Sarma}}, \bibinfo {author} {\bibfnamefont {S.}~\bibnamefont {Adam}},
  \bibinfo {author} {\bibfnamefont {E.}~\bibnamefont {Hwang}}, \ and\ \bibinfo
  {author} {\bibfnamefont {E.}~\bibnamefont {Rossi}},\ }\href@noop {}
  {\bibfield  {journal} {\bibinfo  {journal} {Reviews of modern physics}\
  }\textbf {\bibinfo {volume} {83}},\ \bibinfo {pages} {407} (\bibinfo {year}
  {2011})}\BibitemShut {NoStop}%
\bibitem [{\citenamefont {Bolotin}\ \emph {et~al.}(2009)\citenamefont
  {Bolotin}, \citenamefont {Ghahari}, \citenamefont {Shulman}, \citenamefont
  {Stormer},\ and\ \citenamefont {Kim}}]{bolotin2009observation}%
  \BibitemOpen
  \bibfield  {author} {\bibinfo {author} {\bibfnamefont {K.~I.}\ \bibnamefont
  {Bolotin}}, \bibinfo {author} {\bibfnamefont {F.}~\bibnamefont {Ghahari}},
  \bibinfo {author} {\bibfnamefont {M.~D.}\ \bibnamefont {Shulman}}, \bibinfo
  {author} {\bibfnamefont {H.~L.}\ \bibnamefont {Stormer}}, \ and\ \bibinfo
  {author} {\bibfnamefont {P.}~\bibnamefont {Kim}},\ }\href@noop {} {\bibfield
  {journal} {\bibinfo  {journal} {Nature}\ }\textbf {\bibinfo {volume} {462}},\
  \bibinfo {pages} {196} (\bibinfo {year} {2009})}\BibitemShut {NoStop}%
\bibitem [{\citenamefont {Zhang}\ \emph {et~al.}(2005)\citenamefont {Zhang},
  \citenamefont {Tan}, \citenamefont {Stormer},\ and\ \citenamefont
  {Kim}}]{zhang2005experimental}%
  \BibitemOpen
  \bibfield  {author} {\bibinfo {author} {\bibfnamefont {Y.}~\bibnamefont
  {Zhang}}, \bibinfo {author} {\bibfnamefont {Y.-W.}\ \bibnamefont {Tan}},
  \bibinfo {author} {\bibfnamefont {H.~L.}\ \bibnamefont {Stormer}}, \ and\
  \bibinfo {author} {\bibfnamefont {P.}~\bibnamefont {Kim}},\ }\href@noop {}
  {\bibfield  {journal} {\bibinfo  {journal} {nature}\ }\textbf {\bibinfo
  {volume} {438}},\ \bibinfo {pages} {201} (\bibinfo {year}
  {2005})}\BibitemShut {NoStop}%
\bibitem [{\citenamefont {Goerbig}(2011)}]{goerbig2011electronic}%
  \BibitemOpen
  \bibfield  {author} {\bibinfo {author} {\bibfnamefont {M.}~\bibnamefont
  {Goerbig}},\ }\href@noop {} {\bibfield  {journal} {\bibinfo  {journal}
  {Reviews of Modern Physics}\ }\textbf {\bibinfo {volume} {83}},\ \bibinfo
  {pages} {1193} (\bibinfo {year} {2011})}\BibitemShut {NoStop}%
\bibitem [{\citenamefont {Orlita}\ \emph {et~al.}(2011)\citenamefont {Orlita},
  \citenamefont {Faugeras}, \citenamefont {Grill}, \citenamefont {Wysmolek},
  \citenamefont {Strupinski}, \citenamefont {Berger}, \citenamefont {de~Heer},
  \citenamefont {Martinez},\ and\ \citenamefont
  {Potemski}}]{orlita2011carrier}%
  \BibitemOpen
  \bibfield  {author} {\bibinfo {author} {\bibfnamefont {M.}~\bibnamefont
  {Orlita}}, \bibinfo {author} {\bibfnamefont {C.}~\bibnamefont {Faugeras}},
  \bibinfo {author} {\bibfnamefont {R.}~\bibnamefont {Grill}}, \bibinfo
  {author} {\bibfnamefont {A.}~\bibnamefont {Wysmolek}}, \bibinfo {author}
  {\bibfnamefont {W.}~\bibnamefont {Strupinski}}, \bibinfo {author}
  {\bibfnamefont {C.}~\bibnamefont {Berger}}, \bibinfo {author} {\bibfnamefont
  {W.~A.}\ \bibnamefont {de~Heer}}, \bibinfo {author} {\bibfnamefont
  {G.}~\bibnamefont {Martinez}}, \ and\ \bibinfo {author} {\bibfnamefont
  {M.}~\bibnamefont {Potemski}},\ }\href@noop {} {\bibfield  {journal}
  {\bibinfo  {journal} {Physical review letters}\ }\textbf {\bibinfo {volume}
  {107}},\ \bibinfo {pages} {216603} (\bibinfo {year} {2011})}\BibitemShut
  {NoStop}%
\end{thebibliography}%

\end{document}